\documentclass[12pt]{article}
\usepackage{amsmath}
\usepackage{amsfonts}
\usepackage{amssymb}

\renewcommand{\theequation}{\arabic{section}.\arabic{equation}}

\long\def\symbolfootnote[#1]#2{\begingroup%
\def\thefootnote{\fnsymbol{footnote}}\footnote[#1]{#2}\endgroup}

\usepackage[margin=.8in]{geometry}

\usepackage{mathrsfs}

\def\beq{\begin{equation}}
\def\eeq{\end{equation}}
\def\beqn{\begin{eqnarray}}
\def\eeqn{\end{eqnarray}}

\def\nolabel{\nonumber }

\def\a{\alpha}
\def\b{\beta}
\def\g{\gamma}
\def\G{\Gamma}
\def\d{\delta}

\def\e{\epsilon}
\def\k{\kappa}
\def\L{\Lambda}

\def\om{\omega}
\def\Om{\Omega}

\def\w{\wedge}

\def\Z{\mathcal{Z}}
\def\tM{\widetilde{\mathcal{M}}}

\def\M{\mathscr{M}}
\def\W{\mathscr{W}}

\def\tm{\widetilde{\mathscr{M}}}

\def\K{\mathcal{K}}

\def\V{\mathscr{V}}

\def\ol#1 {\overline#1}
\def\mbf#1 {\mathbf#1}

\def\r#1 {{}^{(#1)}\mathcal{R}}
\def\R#1 {{}^{(#1)}\hat{\mathcal{R}}}

\def\half{\frac{1}{2}}

\def\phe{}

\def\inbar{\,\vrule height1.5ex width.4pt depth0pt}

\def\IC{\relax\hbox{$\inbar\kern-.3em{\rm C}$}}
\def\IQ{\relax\hbox{$\inbar\kern-.3em{\rm Q}$}}
\def\IR{\relax{\rm I\kern-.18em R}}
 \font\cmss=cmss10 \font\cmsss=cmss10 at 7pt
\def\IZ{\relax\ifmmode\mathchoice
 {\hbox{\cmss Z\kern-.4em Z}}{\hbox{\cmss Z\kern-.4em Z}}
 {\lower.9pt\hbox{\cmsss Z\kern-.4em Z}}
 {\lower1.2pt\hbox{\cmsss Z\kern-.4em Z}}\else{\cmss Z\kern-.4em Z}\fi}

\usepackage{hyperref}

\begin{document}

\begin{titlepage}
\setcounter{page}{1}
\vskip 0.5truecm
\rightline{BU-HEPP-06-10}
\rightline{CASPER-06-03}
\rightline{\tt hep-th/0612171}
\rightline{December 2006}
\vskip 0.5truecm
\begin{center}{\huge \sc{The Ricci Curvature}\\
\vskip 0.5truecm
\Huge{of Half-flat Manifolds}}\\
\vskip 0.4truecm
\Large{Tibra Ali\symbolfootnote[1]{Tibra\_Ali@baylor.edu} and Gerald B. Cleaver\symbolfootnote[2]{Gerald\_Cleaver@baylor.edu}}

\large{{\it CASPER, Department of Physics}\\
{\it Baylor University\\One Bear Place \# 97316\\Waco, Texas 76798\\U.S.A.}}
\end{center}
\vskip 0.5truecm
\begin{abstract}
We derive expressions for the Ricci curvature tensor and scalar in terms of intrinsic torsion classes of half-flat manifolds by exploiting the relationship between half-flat manifolds and non-compact $G_2$ holonomy manifolds. Our expressions are tested for Iwasawa and more general nilpotent manifolds. We also derive expressions, in the language of Calabi-Yau moduli spaces, for the torsion classes and the Ricci curvature of the \emph{particular} half-flat manifolds that arise naturally via mirror symmetry in flux compactifications. Using these expressions we then derive a constraint on the K\"ahler moduli space of type II string theories on these half-flat manifolds.
\end{abstract}

\end{titlepage}

\setcounter{footnote}{0}
\section{Introduction}
Flux compactifications (see \cite{grana} and \cite{douglas} for reviews and exhaustive lists of references) have become of central importance in string theory in recent years. The reasons for this are manifold, ranging from a desire to make contact with experiments (in particle physics and cosmology) to the purely theoretical interest (to understand new solutions, geometries and topologies for their own sake). The last few years have seen the gathering of a lot of new `data', yet no synthesis, analogous to the second string revolution, is in sight. In fact, it can be argued that the gathering of mathematical data has just begun, and that this has inspired (and has been inspired by) quite a few breakthroughs in mathematics. Recent years have also seen the maturing of topological string theory (see \cite{neitzke,vonk} for reviews and references) and the birth of the outline of a theory known as topological M-theory \cite{dijkgraaf}, which promises to unify the different topological string theories in a unified framework. Such theories, playing a side role in their infancy, have come more and more to the mainstream discussions of string theory, with many unexpected contributions to the physical side of string theory.
\par
In \cite{vafa}, which is a seminal attempt to bridge topological- and super-string theories, Vafa made the conjecture that mirror symmetry ought to be extendable to the scenario that included compactification of string theory on Calabi-Yau manifolds with fluxes. This has been verified in the low-energy limit for different cases. For example in \cite{louis} it was shown that type IIA and type IIB theories with RR fluxes continue to be mirror symmetric if they are compactified on a pair of mirror Calabi-Yau manifolds, $\M$ and $\W$. Recall that the RR field strengths are even forms in IIA and odd forms in IIB, and so the extension of mirror symmetry with RR fluxes is consistent, although by no means trivially so, with the intuition of mirror symmetry exchanging odd and even cohomologies in the Hodge diamonds of the mirror pair. Since the NS field strength is a three form and is the same in both type II theories, the results of \cite{louis} raises the obvious question of what is the mirror dual of, say, type IIB theory on a Calabi-Yau $\W$ with just NS fluxes turned on. Vafa had suggested that the dual of IIB on $\W$ with NS fluxes also has fluxes but now these fluxes are `even NS fluxes' that are encoded in the dilaton and the graviton (which are also in the NS sector). In other words, Vafa suggested that the dual ought to be an almost complex manifold whose complex structure would \emph{not} be integrable. This lack of integrability would encode the missing NS fluxes, without recourse to the usual `matter sector', say, as in \cite{louis}.
\par
Vafa's intuition was made concrete in the astonishing work \cite{gurrieri} which showed that the relevant geometry dual to that of Calabi-Yau $\W$ with NS fluxes, are a special class of \emph{half-flat} manifolds which we shall denote by $\tm$ in this paper. These manifolds are, in a sense, deformations of the Calabi-Yau manifold $\M$ which is the canonical mirror dual of $\W$.\footnote{By `canonical mirror dual' we mean `mirror dual in the absence of fluxes.'}  \emph{Generic} half-flat manifolds, $\tM$, are a special class of $SU(3)$ manifolds which are the minimal ingredient for the $\mathcal{N}=2$ reduction of ten-dimensional type II string theory \cite{minasian}. 
\par
Half-flat manifolds had already made their appearance in the work of Hitchin \cite{hitchin}, and they play an important role in the construction of a special type of $G_2$ holonomy manifold. It is interesting to note that Hitchin's work emphasizes the form-aspects of the special geometries in six and seven dimensions, and thus is very close in spirit to topological string/M theories and the related form-theories of gravity.
\par
In this paper we were motivated by taking a closer look at the metric sector of the half-flat manifolds $\tm$ proposed in \cite{gurrieri}, in particular we wanted to know if the Ricci tensor of $\tm$ plays any role in the low energy effective theory. At the time of the onset of this work there were no formulae for the Ricci curvature of half-flat manifolds available in the literature, and we decided to derive expressions for it. Recently, as our work of the `physics' part was nearing completion, which is based on the mathematical results that we derived earlier, we learned in \cite{minasian} of a paper \cite{bedulli} in the mathematics archives which computes the Ricci curvature of  manifolds with $SU(3)$ structure. Thus the first half of our paper will have some overlap with \cite{bedulli} in content but not in technique, language or point of view.
\par
\emph{The central results of our paper are the following: We have derived expressions, eqs. (\ref{eq:Ricci}) and (\ref{eq:scalar}), for the Ricci curvature of generic half-flat manifolds in terms of their intrinsic torsion by thinking of them as hypersurfaces that foliate a non-compact $G_2$ holonomy cylinder \emph{\`a la} Hitchin \cite{hitchin}. We then derive the Ricci curvature, eqs. (\ref{eq:mixed}), (\ref{eq:pure}) and (\ref{eq:ricciscalar}), of the particular class of half-flat manifolds $\tm$ that arise from mirror symmetry \cite{gurrieri}.  We then use a consistency argument involving the Ricci curvature of $\tm$ in $\mathcal{L}_{\mathrm{eff}}$ to derive a condition, eqs. (\ref{eq:infcond}) or (\ref{eq:intcond}), on the K\"ahler moduli space. This leads, by using one of Hitchin's flow equations, to a formula (\ref{eq:volflow}) for the flow of the volume of $\tm$.}
\par
At the risk of repeating ourselves, we now give a brief overview of the different sections of our paper. In section two we review properties of six-dimensional manifolds with $SU(3)$ structure, considering in particular those of Calabi-Yau and half-flat manifolds. Intrinsic torsion classes of the manifolds, and expression of the exterior derivatives of the fundamental two form $J$ and three form $\Omega$ in terms of them, are discussed. In section three we show that the intrinsic torsion of half-flat manifolds is equivalent to their extrinsic curvature when embedded inside seven dimensional manifolds of $G_2$ holonomy. Then in section four, by exploiting the relationship between half-flat manifolds with $SU(3)$ structure and non-compact manifolds with $G_2$ holonomy, we derive expressions for the Ricci curvature tensor and scalar for half-flat manifolds in terms of the torsion classes. In section five we verify our results of section four by comparing Ricci curvature computed using torsion classes to that determined from standard affine connection expressions. We make our comparisons for a subset of half-flat manifolds known as nilmanifolds (which are derived from nilpotent Lie algebras and include Iwasawa manifolds as examples). Then in section six we derive expressions for the torsion classes of the \emph{particular} half-flat manifolds $\tm$ in the language of the moduli spaces of the Calabi-Yau manifold $\M$, and then derive expressions for the Ricci curvature for $\tm$. We then turn our attention to a particular term in the low energy effective action (derived in \cite{gurrieri}) and study its variation. We derive two expressions for this variation which from consistency arguments must be the same, and from that we derive a condition on the K\"ahler moduli space of the compactification.  We show that this condition reduces to a trivial identity in the limiting case when the `underlying' Calabi-Yau $\M$ has a one dimensional K\"ahler moduli space. In section seven we conclude the body of our paper with a discussion of our findings and some comments on further work that needs to be done.
\par
Some background material for the paper is presented in the appendices. In Appendix \ref{ap:difforms} we present a review of differential forms and our conventions. In Appendix \ref{ap:cymod} we review the K\"ahler deformations, complex structure deformations, and the large structure limit associated with Calabi-Yau manifolds. In Appendix \ref{ap:details} we give details of the derivations of the results presented in Section \ref{sec:defCY}.
\setcounter{equation}{0} 
%
\section{Manifolds with $SU(3)$ structure: Calabi-Yau and Half-flat}\label{sec:su3}
A six dimensional orientable manifold $M$ is said to have $SU(3)$ structure if it admits \emph{only} one no-where vanishing Majorana spinor $\eta$. It can be shown that the existence of such a spinor implies that the structure group of the frame bundle of $M$ is reduced from $SO(6)$ to $SU(3)$. This globally defined spinor is the singlet in the decomposition of the $\bf{4}$ of $SU(4)$ (which is the spin cover of $SO(6)$) under $SU(3)$:
\begin{equation}
\bf{4}\rightarrow \bf{3} + \bf{1}.
\end{equation}
Another way of defining $SU(3)$ structure is to say that $M$ admits a non-degenerate almost hermitian structure $J_m{}^n$ (which is not necessarily integrable) and a globally defined 3-form $\Omega_{mnp}$ which is of type $(3,0)$ with respect to $J_m{}^n$. It can be shown that such objects define a Riemannian structure $g_{mn}$ on $M$ \cite{joyce}.
\par
Given the Riemannian structure $g_{mn}$ one can construct a $(1,1)$-type 2-form, which we denote by the same symbol $J$, via
\begin{equation}
J_{mn} =J_m{}^p g_{pn}.
\end{equation}
The quantities $J_{mn}$  and $\Om_{mnp}$ can be constructed from the globally defined spinor $\eta$ in the following way
\begin{equation}
\begin{split}
J_{mn} &=i\bar{\eta}\G_7\G_{mn}\eta \\
\Om^+_{mnp} &= -\bar{\eta} \G_7 \G_{mnp}\eta\\
\Om^-_{mnp} &= i \eta\G_{mnp} \eta\\
\Om &= \Om^+ + i \Om^-
\end{split}
\end{equation}
where $\G_{m_1\dots m_n}$ are the antisymmetrized products of the $n$ gamma matrices and $\G_7$ is the six-dimensional chirality operator. Our convention for the gamma matrices is that they are all imaginary (including $\G_7$) and hermitian. Conjugate spinors are given by $\bar{\psi} = \psi^\dag$, and the Majorana condition is $\bar{\psi} = \psi^T$. Also we denote by $\Om^+$ and $\Om^-$ the real and the imaginary part of $\Om$, respectively. We have taken $\eta$, the globally defined  eight-component Majorana spinor on $M$, to be commuting.
\par
This construction is of course local and for the manifold to have $SU(3)$ structure the spinor $\eta$ must remain a singlet under parallel propagation around closed loops via some connection $\nabla'$. I.e.
\begin{equation}
\nabla' \eta =0. \label{eq:parallel}
\end{equation}
However, this connection $\nabla'$ may not be the torsion-free unique Levi-Civita connection, which we denote by $\nabla$, that is constructed from $g_{mn}$. If it is, then the integrability condition of eq.(\ref{eq:parallel}) implies that holonomy group of the Levi-Civita connection, which we denote by $\mathscr{H}(\nabla)$, is $SU(3)$, and the corresponding spinor $\eta$ is said to be covariantly constant or just constant. Such manifolds are K\"ahler with vanishing first Chern class, and the corresponding metric is Ricci-flat. These manifolds are the Calabi-Yau manifolds.
\par
If on the other hand $\eta$ is not constant then the holonomy group of the Levi-Civita connection will not be $SU(3)$. Given the Levi-Civita connection and
\begin{equation}
\nabla \eta\ne 0
\end{equation}
we can always construct a connection $\nabla'$ such that (\ref{eq:parallel}) is satisfied \cite{ffs}. The difference between these two connections is given by the contorsion $\kappa'$ or equivalently the torsion $W'$, which lies in $\Lambda^1\otimes \Lambda^2\cong \Lambda^1 \otimes \mathfrak{so}(6)$. (We use the convention that $\mathfrak{g}$ denotes the Lie algebra of the Lie group $G$). However, when structure group is reduced to $SU(3)$, the difference between the two connections is encoded in the intrinsic contorsion $\kappa$ or the intrinsic torsion\footnote{Torsion and contorsion are isomorphic and when there is no room for confusion we use the terms interchangeably.} $W$, which is actually in $\Lambda^1 \otimes \mathfrak{su}(3)^\perp$, where $\mathfrak{su}(3)^\perp$ is defined by the decomposition $\mathfrak{so}(6)=\mathfrak{su}(3)\oplus \mathfrak{su}(3)^\perp$.
Thus we have from eq.(\ref{eq:parallel})
\begin{equation}
\nabla\eta+\kappa\eta=0
\end{equation}
 and intrinsic torsion encodes how $\mathscr{H}(\nabla)$ deviates from $SU(3)$.
\par
Since $\Lambda^1\cong \bf{3} \oplus \bf{3}$ and $ \mathfrak{su}(3)^\perp = \bf{1}\oplus \bf{3} \oplus \bf{\bar{3}}$ under $SU(3)$, the intrinsic torsion breaks up into various $SU(3)$ modules, and this provides a useful scheme for the classification of manifolds with $SU(3)$ structure\footnote{From now on all representations shall be representations of $SU(3)$ unless otherwise indicated.}. It can be shown that the intrinsic torsion for $M$ breaks up into five $SU(3)$ modules \cite{chiossi}
\begin{equation}\begin{array}{ccccccccc}
(\mathbf{1} \oplus \mathbf{1}) &\oplus & (\mathbf{8} \oplus \mathbf{8}) &\oplus & (\mathbf{6} \oplus \mathbf{\bar{6}}) & \oplus  & (\mathbf{3} \oplus \mathbf{\bar{3}}) &\oplus & (\mathbf{3'} \oplus \mathbf{\bar{3}'})\\
W_1 &\oplus & W_2 &\oplus & W_3 &\oplus & W_4 &\oplus & W_5 
\end{array}
\end{equation}
where these modules are called \emph{torsion classes} and denoted by $W_i$. Thus the class of $SU(3)$ manifolds known as Calabi-Yau manifolds may be characterized by $W_i =0$ $\forall i$.
\par
It should be clear from the foregoing discussion that in the general case of $SU(3)$ structure, neither $J$ (the 2-form corresponding to $J_m{}^n$) nor $\Om$ would be closed. $dJ$ and $d\Om$, which are in the $\mathbf{20}$ and $\mathbf{24}$ representations of $SO(6)$, respectively, decompose, under  $SU(3)$, in the following way \cite{chiossi}
\begin{equation}
\begin{array}{cccccc}\\
dJ= & -\frac{3}{2} (W_1 \ol{\Om})^- & + & W_4 \w J & + &W_3\\
\mathbf{20} =&(\mathbf{1}\oplus \mathbf{1}) &\oplus &  (\mathbf{3}\oplus \mathbf{\bar{3}}) & \oplus  & (\mathbf{6}\oplus \mathbf{\bar{6}})
\end{array} \label{eq:dj}
\end{equation}
and
\begin{equation}\begin{array}{cccccc}
d\Om = &W_1 J\w J &+& W_2 \w J &+& \ol{W}_5 \w \ol{\Om}\\
\mathbf{24}=&(\mathbf{1}\oplus \mathbf{1}) &\oplus  & (\mathbf{8}\oplus \mathbf{\bar{8}}) & \oplus & (\mathbf{3'}\oplus \mathbf{\bar{3}'}). 
\end{array} \label{eq:dom}
\end{equation}
\par
As explained in the introduction, we are interested in a class of manifolds with $SU(3)$ structure which arise from considering mirror symmetry on compactifications of string theories on Calabi-Yau manifolds with NS fluxes. Such manifolds are known as \emph{half-flat} manifolds and they are characterized by\footnote{Note that $\Om$ is defined only up to multiplication by a phase, hence there are two natural options for (\ref{eq:conditions1}). We choose our convention to be consistent with \cite{gurrieri} which differs from that of \cite{chiossi} by a factor of $-i$. \label{footnote:convention}}
\begin{equation}
W^-_1=W^-_2=W_4=W_5=0 \label{eq:conditions1}
\end{equation} 
where by $W^{+}_i$ ($W^{-}_i$) we denote the real (imaginary) part of $W_i$. In terms of $J$ and $\Om$ the requirement of half-flatness (\ref{eq:conditions1}) becomes \cite{chiossi,gurrieri}
\begin{equation}
\begin{split}
J\w d J &=0 \\
d \Om^- & =0.
\end{split} \label{eq:conditions2}
\end{equation}
\setcounter{equation}{0}
%
\section{Half-flat and $G_2$ Holonomy Manifolds:\\ Intrinsic Torsion = Extrinsic Curvature}\label{sec:halfflat}
In this section we demonstrate, by constructing the explicit maps, that the intrinsic torsion of half-flat manifolds is equivalent to the extrinsic curvature of the embedding of such manifolds inside a manifold with $G_2$ holonomy. 
\par
Let $\Z$ be a seven-dimensional Riemannian manifold with $G_2$ holonomy. This means that $\Z$ comes equipped with a 3-form $\phi$ and a 4-form $\tilde{\phi}$, both of which are globally defined, and invariant under a $G_2$ transformation of the tangent bundle; and that both $\phi$ and $\tilde{\phi}$ are closed. Then, they induce a Riemannian structure $\hat{g}_{MN}$ on $\Z$ such that
\begin{equation}
\tilde{\phi} =\hat{*}\phi
\end{equation}
where the Hodge star operator $\hat{*}$ is defined with respect to $\hat{g}_{MN}$. These conditions are equivalent to the apparently more stringent condition
\begin{equation}
\hat{\nabla}\phi =0
\end{equation}
where $\hat{\nabla}$ is the Levi-Civita connection constructed from $\hat{g}_{MN}$. $\phi$ and $\tilde{\phi}$ are called the associative and co-associative $G_2$-forms, respectively, and these labels derive from the fact that the components of $\phi$ are the structure constants of the automorphism group of the imaginary octonians.
\par
Another way of saying that $\Z$ has $G_2$ holonomy is to say that there is only one covariantly constant Majorana spinor $\eta$,
\begin{equation}
\hat{\nabla}\eta=0, \label{eq:G2parallel}
\end{equation}
on $\Z$. Then this spinor can be used to define the associative and the co-associative forms via
\begin{equation}\begin{split}
\phi_{MNP} & =i\bar{\eta}\G_{MNP}\eta\\
\tilde{\phi}_{MNPQ}&=\bar{\eta}\G_{MNPQ}\eta
\end{split}
\end{equation}
\par
A seven dimensional manifold with $G_2$ holonomy defines locally an $SU(3)$ structure on a six dimensional sub-manifold. As mentioned in the introduction we want to exploit the relationship between half-flat manifolds $\tM$ with $SU(3)$ structure and non-compact $G_2$ holonomy manifolds $\Z$. Such a $G_2$ structure has been called `dynamic' in \cite{chiossi} to reflect the fact that the global embedding of $SU(3)$ into $G_2$ holonomy reduces to a Hamiltonian mechanical system \cite{hitchin}. To make this connection between seven and six dimensions, we dress the discussion on the intrinsic torsion of $\tM$ in terms of how $\tM$ is embedded in $\Z$.
\par
The class of $G_2$ holonomy manifolds that we are interested in has the useful property that their Ricci-flat metric can be globally written as \footnote{Note that this is quite different from saying that this is what the metric looks like in Gaussian local coordinates.}
\begin{equation}
\begin{split}
d\hat{s}^2 &\equiv\hat{g}_{MN} dy^M dy^N\\
&= dz^2 + g_{mn}(z,y) dy^m dy^n. \label{eq:G2metric}
\end{split}
\end{equation}
Of course this does not mean that $\{y^m\}$ are global coordinates. Manifolds which admit metrics of the above form, regardless of their holonomy, have been called `generalized cylinders' in \cite{bar}. $g_{mn}(z,y)$ denote the metric on the half-flat leaves ${}^z\tM$ that foliate $\Z$, which is induced from the `dynamical' embedding of the $SU(3)$ structure within the $G_2$ structure . We should adopt the notation, consistent with the last section, that ${}^z \nabla$ denotes the Levi-Civita connection on ${}^z \tM$ for the metric $g_{mn}(z,y)$; however, from now on we omit the label $z$ on the quantities on the the half-flat manifold, unless there is a cause for confusion. Then via the Gau\ss-Weingarten equation we get from eq.(\ref{eq:G2parallel}) the following equation for the Levi-Civita connection of $\tM$
\begin{equation}
\nabla_m \eta +\frac{1}{2} K_m{}^n \G_n \G^z \eta=0 \label{eq:G2Killing}
\end{equation}
where $K_{mn}=1/2 \; \partial_z g_{mn}$ is the second fundamental form of the embedding of $\tM$ in $\Z$.  The $SU(3)$ structure of $\tM$ is then given in terms of the $G_2$ structure by
\begin{equation}
\begin{split}
J_{mn} &=\phi_{z mn}\\
\Om^-_{mnp} &= \phi_{mnp} \\
 \Om^+_{mnp} &= \tilde{\phi}_{mnpz}.
\end{split}
\end{equation}
On a $G_2$ holonomy manifold we have
\begin{equation}
\bar{d}\phi=\bar{d}\hat{*} \phi=0 \label{eq:closure}
\end{equation}
where we have used $\bar{d}$ to denote the seven dimensional exterior derivative operator. Denoting by $d$ the exterior derivative operator on $\tM$ and setting
\begin{equation}
\bar{d}=dz\frac{\partial}{\partial z} + d
\end{equation}
we find that (\ref{eq:closure}) leads to the half-flat conditions (\ref{eq:conditions2}) in addition to the compatibility conditions,
\begin{eqnarray}
dJ &=&\frac{\partial \Om^-}{\partial z} \label{eq:flow1}\\
d\Om^+ &=&-J\w \frac{\partial J}{\partial z}, \label{eq:flow2}
\end{eqnarray}
which lead to the seven dimensional $G_2$ holonomy metric above \cite{hitchin,chiossi}. Eqs.\ (\ref{eq:flow1}) and (\ref{eq:flow2}) are known as Hitchin's flow equations.
\par
Note that on a general six dimensional manifold with $SU(3)$ structure the intrinsic torsion has 42 independent components which the half-flatness condition reduces by half. Also, note that, for a general seven dimensional metric of the form (\ref{eq:G2metric}) the second fundamental form has exactly 21 independent components. As is obvious from (\ref{eq:G2Killing}) then, there is a one-to-one map between the intrinsic torsion of a half-flat manifold and the second fundamental form of its embedding into a $G_2$ holonomy manifold. In fact it is easy to show that the intrinsic contorsion $\kappa$ mentioned in the previous section is related to the extrinsic curvature $K$ via
\begin{equation}
\kappa_{rst} = \frac{1}{2}K_r^m \Om^+_{mst}. \label{eq:intrncon}
\end{equation}
\par
We now construct the map between the individual torsion classes and the extrinsic curvature. This map can be found by using (\ref{eq:G2Killing}) to calculate the exterior derivatives (\ref{eq:dj}) and (\ref{eq:dom}). The relevant expressions to start from are the derivatives of $J$ and $\Om$ in terms of $K_{mn}$ 
\begin{equation}
\begin{split}
\nabla_m \Om^+_{pqr}&= K_m^n \left( J_{np} J_{qr} - J_{nq} J_{pr} + J_{nr} J_{pq}\right)\\
\nabla_m \Om^-_{pqr} &= -K_{mr} J_{pq}+ K_{mq} J_{pr} - K_{mp} J_{qr}\\
\nabla_m J_{pq} &= K_m^n \Om^-_{npq},
\end{split}
\end{equation}
and then anti-symmetrizing we get
\begin{eqnarray}
(dJ)_{mpq} &=& 3 K_{[m}^n \Om^-_{|n|pq]} \label{eq:1st}\\
(d\Om)_{mpqr} &=&  12 K_{[m}^n J_{|n|p} J_{qr]} \label{eq:2nd}.
\end{eqnarray}
Using these expressions it is easy to show that the intrinsic contorsion (\ref{eq:intrncon}) is equivalent to
\begin{equation}
\kappa_{rst}=-\frac{1}{2} \left(\nabla_r J_s{}^q\right) J_{qt}.
\end{equation}
This is the contorsion used in analyzing the geometry of almost hermitian manifolds in \cite{ffs} and was introduced in \cite{yano}.
\par
To obtain the torsion classes from eqs.\ (\ref{eq:1st}, \ref{eq:2nd}), we proceed in the following fashion (see \cite{chiossi} for more details). By contracting $\overline{\Om}$ or $J\w J$ with $dJ$ or $d\Om$, respectively, one derives $W_1$.  To obtain then $W_5$, one contracts $\Om$ with $d\Om$. Subtracting these two from $d\Om$ one obtains $W_2$. The primitive (2,1) part of $dJ$ gives $W_3$ and its orthogonal (2,1) complement gives $W_4$. This way we find
\begin{equation}
\begin{split}
W^+_1 &=\frac{1}{3}K\\ 
(W^+_2)_{mn} &= -\frac{1}{3} K J_{mn} + K_m^r J_{rn} - K_n^r J_{rm}\\
(W_3)_{mnp} &= -\frac{3i}{2} \Pi^-_{[m}{}^r K_{|r|}^q \Om_{np]q}
\end{split} \label{eq:torclass}
\end{equation}
as well as $W^-_1=W^-_2=W_4=W_5=0$. Note that $W^+_2$ is a primitive (1,1)-form and $W_3$ is a primitive (2,1)-form. In the above equations $K$ denotes the trace of the second fundamental form. Also we have the usual projection operators
\begin{equation}
\Pi^\pm_m {}^n = \frac{1}{2} \left( \d^n_m \mp i J_m{}^n\right),
\end{equation}
which decompose the tangent space into (1,0)- and (0,1)-types, and satisfy the usual relationships for projection operators,
\begin{equation}
\begin{split}
\Pi^+ + \Pi^- &= 1\\
\Pi^\pm \Pi^\pm &=\Pi^\pm \\
\Pi^\mp \Pi^\pm =\Pi^\pm \Pi^\mp &=0.
\end{split}
\end{equation}
 Eqs.\ (\ref{eq:torclass}) can be easily inverted
\begin{equation}
\begin{split}
K_{mn} &= \left(\Pi^+_m{}^{m' }+\Pi^-_m{}^{m' }\right)\left(\Pi^+_n{}^{n' }+\Pi^-_n{}^{n' }\right)K_{m'n'}\\
&=  \frac{1}{2} W^+_1 g_{mn} +\frac{1}{2} \left(W^+_2\right)_{mp} J_n{}^p - \frac{1}{4} \left[ \left(W_3\right)_{mpq} \ol{\Om}_n {}^{pq}\right]^-
\end{split}\label{eq:ktorclass}
\end{equation}
\par
In the following pages we shall implicitly use contorsion in our calculations and it is convenient to have the relationship between the contorsion components and the intrinsic torsion:
\begin{equation}\begin{split}
\kappa_{srt}=& \frac{1}{4} W^+_1 \Om^+_{srt} + \frac{1}{4} (W_2^+)_{sq}\Om^-{}^q{}_{rt} \\
&+\frac{i}{2} \left\{(W_3)_{suv}\Pi^+_r {}^u \Pi^+_t {}^v - (\ol{W}_3)_{suv}\Pi^-_r {}^u \Pi^-_t {}^v\right\}.
\end{split}
\end{equation}
The contorsion is anti-symmetric in its last two indices. For half-flat manifolds, however, the contorsion is completely traceless
\begin{equation}
\kappa_{sr}{}^s= \kappa_s{}^s{}_r=0.
\end{equation}
This follows easily from the expression given above when one uses the facts that $W^+_2$ and $W_3$ are primitive.
\setcounter{equation}{0}
%
%
\section{The Ricci Curvature of Half-flat Manifold}\label{sec:ricci}

In this section we give expressions for the Ricci curvature in terms of the torsion classes. However, before we do so let us state our convention for the various curvature tensors. The Riemann curvature tensor is defined by
\begin{equation}
\left[\nabla_m,\nabla_n\right] V^p = R_{mnq}{}^p V^q \label{rctdef}
\end{equation}
and the Ricci tensor and scalar are 
\beqn
R_{mn} = R_{mpn}{}^p \label{riccidef}
\eeqn and 
\beqn
R=g^{mn}R_{mn}. \label{scalardef}
\eeqn
 These definitions are valid for both six and seven dimensions.
\par
Next, using the fact that $\Z$ is Ricci-flat and the Gau\ss-Codazzi relations it is easy to show that the Ricci curvature of $\tM$ is given by
\begin{equation}\begin{split}
R_{mn} &= -\frac{\partial K_{mn}}{\partial z} - K K_{mn} +2 K_{mr} K_{n}^r. \label{eq:RicciK}
\end{split}
\end{equation}
Because of $G_2$ holonomy the Riemann curvature tensor of $\Z$ satisfies the following `self-duality' relationship \cite{acharya}:
\begin{equation}
\hat{R}_{MNPQ} = \frac{1}{2}\tilde{\phi}_{PQ}{}^{RS} \hat{R}_{MNRS}.
\end{equation}
Taking $\{M=m, N=z, P=p, Q=z\}$ components of this equation, we can derive an expression for $\partial_z K_{mn}$ which can then be used in (\ref{eq:RicciK}) to obtain the following expression for the Ricci curvature of $\tM$:
\begin{equation}
R_{mn} = \Om^+_n{}^{rs} \nabla_r K_{sm} - K K_{mn} + K_{mr} K^r_n.
\end{equation}
We can now use the expressions for the second fundamental form derived in the previous section to express the Ricci curvature in terms of torsion classes:
\beqn
\begin{split}
R_{mn}=& -\frac{3}{4}\Om^-_n{}^{ps}\partial_{[p} W^+_2{}_{sm]}-\frac{1}{2}J_{(m}{}^p\nabla^s\left[W_3{}_{|sp|n)} + \overline{W}_3{}_{|sp|n)} \right]\\
&\phe -\frac{5}{4}g_{mn} (W_1^+)^2 - \frac{1}{4} W_1^+  \; W^+_2{}_{mr} J_n{}^r -\frac{i}{8} W_1^+\left[(W_3\cdot \overline{\Om})_{mn} -(\overline{W}_3\cdot {\Om})_{mn} \right]\\
&\phe  -\frac{1}{2} W_2^+{}_{mq}\; W_2^+{}_n{}^q -\frac{1}{16}W_2^+{}_m{}^r \left[(W_3\cdot \overline{\Om})_{nr} +(\overline{W}_3\cdot {\Om})_{nr} \right]\\
&\phe -\frac{1}{8}W_2^+{}_v {}^r \left[W_3{}_{mu}{}^v \overline{\Om}_n{}^u{}_r +\overline{W}_3{}_{mu}{}^v {\Om}_n{}^u{}_r\right]\\
&\phe +\frac{1}{4}\left[W_3{}_{muv} {\overline{W}}_3{}_n{}^{st} +W_3{}_{nuv} {\overline{W}}_3{}_m{}^{st}\right] \Pi^+_s {}^u \Pi^+_t {}^v \\
&\phe - \frac{1}{2}\left[W_3{}^s{}_{mu} W_3{}^u {}_{pq} \Pi^+_n {}^p \Pi^+_s {}^q +\overline{W}_3{}^s{}_{mu} \overline{W}_3{}^u {}_{pq} \Pi^-_n {}^p \Pi^-_s {}^q \right]\\
&\phe +\frac{1}{2} \left[ W_3{}_{suv} \Pi^+_{(m} {}^u \Pi^+_{|q|} {}^v \ol{W}_3^{sq}{}_{n)} +\ol{W}_3{}_{suv} \Pi^-_{(m} {}^u \Pi^-_{|q|} {}^v W_3^{sq}{}_{n)}\right]  \label{eq:Ricci}
\end{split}
\eeqn
where we have used the notation $(W_3\cdot \ol{\Om})_{mn} \equiv W_3{}_{mpq} \ol{\Om}_n{}^{pq}$, etc. The scalar curvature is
\begin{equation}
R = \frac{15}{2} (W_1^+)^2- \frac{1}{4} W^+_2{}_{mn} W_2^+{}^{mn} -\frac{1}{6} W_3{}_{pqr} \ol{W}_3{}^{pqr}. \label{eq:scalar}
\end{equation}
In the above calculations we have found the following $SU(3)$ identities to be of great use \cite{lust}:
\begin{equation}\begin{split}
\ol{\Om}^{pqr}\Om_{stu} &= 48 \Pi^+_s {}^{[p} \Pi^+_t {}^q  \Pi^+_u {}^{r]}\\
\ol{\Om}^{pqr}\Om_{pst} &= 16 \Pi^+_s {}^{[q} \Pi^+_t {}^{r]}\\
\ol{\Om}^{pqr}\Om_{pqs} &= 16 \Pi^+_s {}^{r}. 
\end{split}\label{eq:projident}
\end{equation}
The expression (\ref{eq:Ricci}) for the Ricci curvature in terms of the torsion classes is the main technical result of this paper. In the rest of the paper we study some applications of eqs.\ (\ref{eq:Ricci}, \ref{eq:scalar}).
\setcounter{equation}{0}
%
\section{Iwasawa and General Nilpotent Manifolds}\label{sec:nil}
Our computation of the Ricci curvature in the previous section is algebraically quite involved. Thus we think it is a good idea to have some checks on the expressions (\ref{eq:Ricci},\ref{eq:scalar}) from a class of manifolds whose Ricci curvatures are already known. To this end we turn to nilmanifolds in this section.
\par
Half-flat manifolds can be nilmanifolds. This class includes the Iwasawa manifolds, which are compact quotient spaces of the form $M=\Gamma\backslash G$, where $G$ is the complex Heisenberg group and $\Gamma$ is the the subgroup of Gaussian integers \cite{lopes}.  The points defined by the complex coordinates $z_{i=1,2,3}$ and $z^{'}_{i=1,2,3}$ are identified on the Iwasawa manifold whenever
\beqn
	\left(\begin{array}{ccc}
						1&z^{'}_{1}&z^{'}_{3}\\
						0&1        &z^{'}_{2}\\
						0&0        &1\end{array}
	\right)
	=	
	\left(\begin{array}{ccc}
						1&m_{1}    &m_{3}\\
						0&1        &m_{2}\\
						0&0        &1\end{array}
	\right)	
	\left(\begin{array}{ccc}
						1&z_{1}    &z_{3}\\
						0&1        &z_{2}\\
						0&0        &1\end{array}
	\right),
\label{Iwamandef}
\eeqn
where $m_{i=1,2,3}\in \IZ + i \IZ$ \cite{lopes,candelas4}. Thus,
\beqn
z_1 &\sim& z_1 + m_1 \nolabel\\
z_2 &\sim& z_2 + m_2 \label{iwamat}\\
z_3 &\sim& z_3 + m_3 + m_1 z_2
\eeqn  
Such manifolds can be interpreted as twisted tori.

The Iwasawa manifolds and more general nilmanifolds provide a means of explicit verification of our torsion class expressions for the Ricci tensor of the affine connection, (\ref{eq:Ricci}), and the corresponding Ricci scalar, (\ref{eq:scalar}) (which is derived independently of (\ref{eq:Ricci})).
The Ricci tensor $R_{m n}$ for a particular manifold can, of course, be determined directly from the affine (Christoffel) connection $\Gamma$,
\beqn
R_{m n}&\equiv& R^{r}_{m r n} = g^{r s}R_{s m r n} \label{Rtx}\\
       &=&      \Gamma^{r}_{m r,n} -\Gamma^{r}_{m n,r }
                +\Gamma^{s}_{m r} \Gamma^{r}_{n s} 
                -\Gamma^{s}_{m n} \Gamma^{r}_{r s}, \label{RtxL}
\eeqn
where the affine connection is defined by,
\beqn
\Gamma^{s}_{m n} \equiv \half g^{r s}
                   \left[g_{n r ,m} + g_{m r,n} - g_{n m ,r}\right].
\label{affinedef}
\eeqn
Thus, our torsion class expression for the Ricci tensor, (\ref{eq:Ricci}), and for the Ricci scalar, (\ref{eq:scalar}), derived using the Gau\ss-Codazzi relations and the $G_2$ holonomy `self-duality' relationship
can be compared, respectively, to affine expression (\ref{RtxL}) and its trace.
In order to substantiate the veracity of different subsets of terms in (\ref{eq:Ricci}) and (\ref{eq:scalar}),  
we performed this comparison for three different manifolds.
Two of our three chosen manifolds are Iwasawa manifolds \cite{lust,lopes} and our third is a more complicated, two-step example of a nilmanifold \cite{chiossi}, produced by slight modification to the first Iwasawa manifold. 
 
The outcomes of our comparisons were positive. The validity of (\ref{eq:Ricci}) and (\ref{eq:scalar}) was supported: For the first of the Iwasawa manifolds, only $W^+_1$ and $W^+_2$ are non-zero \cite{lust}; for the second, only $W_3$ is non-zero \cite{lopes}; for the more general nilpotent manifold, $W^+_1$, $W^+_2$, and $W_3$ are all non-zero \cite{chiossi}. Thus, we separately verified the contributions from $\{W^+_1,\, W^+_2 \}$, from $\{W_3\}$, and the mixed contributions from the two subsets. 

The metric for Iwasawa manifolds can perhaps be most simply expressed in terms of an orthonormal tangent basis of one-forms $e^i$, $i= 1,\, ...\, ,\, 6$. 
In this basis, the fundamental 2-form $J$ and the 3-form $\Omega$ may be expressed locally as, 
\beqn
 J     &=& e^1\wedge e^2 + e^3\wedge e^4 + e^5\wedge e^6. \label{kform}\\
\Omega &=& \left[e^1\wedge e^3\wedge e^6 + e^1\wedge e^4\wedge e^5 + e^2\wedge e^3\wedge e^5 
           - e^2\wedge e^4\wedge e^6 \right]
\nolabel\\
       && -i \left[e^1\wedge e^3\wedge e^5 - e^1\wedge e^4\wedge e^6 - e^2\wedge e^3\wedge e^6 - e^2\wedge e^4\wedge e^6 \right]. 
\label{3form}   
\eeqn
(Recall footnote 4 regarding the phase of $\Omega$.)

The $\{W^+_1,\, W^+_2 \}$ Iwasawa manifold of \cite{lust} is specified by the metric
\beqn
ds^2 = \sum_{i=1}^4 (dx^i)^2 + (dx^5 - x^1 dx^4 + x^3 dx^2)^2 + (dx^6 - x^1 dx^3 - x^4 dx^2)^2.
\label{ds2w12}
\eeqn
The coordinates $x^i$ can be used to define the following tangent basis
\beqn   
e^j &=& dx^j,\, j= 1,\, ...\, ,\, 4,\label{I12ej}\\
e^5 &=& dx^5 - x^1 dx^4 + x^3 dx^2, \label{I12e5}\\
e^6 &=& dx^6 - x^1 dx^3 - x^4 dx^2, \label{I12e6}
\eeqn
with exterior derivatives of the tangent forms specified by
\beqn   
de^j &=& 0,\, j= 1,\, ...\, ,\, 4,\label{I12dej}\\
de^5 &=& -e^1 \wedge e^4 - e^2 \wedge e^3, \label{I12de5}\\
de^6 &=& -e^1 \wedge e^3 + e^2 \wedge e^4, \label{I12de6}
\eeqn

The torsion classes, expressed in the tangent basis, are
\beqn
W^+_1 &=& -\frac{2}{3} \label{I12W1}\\ 
W^+_2 &=& -\frac{4}{3} \left[ e^1 \wedge e^2 + e^3 \wedge e^4 - 2 e^5 \wedge e^6 \right]. \label{I12W2}\\
W_3   &=& 0 \label{I12W3}
\eeqn
With only non-zero $W^+_1$ and $W^+_2$, (\ref{eq:Ricci}) reduces to 
\begin{equation}\begin{split}
R_{mn}(W^+_1,W^+_2) &= -\frac{3}{4}\Om^-_n{}^{ps}\partial_{[p} W^+_2{}_{sm]} -\frac{5}{4}g_{mn} (W_1^+)^2 \\
&- \frac{1}{4} W_1^+ \; W^+_2{}_{mr} J_n{}^r
-\frac{1}{2} W_2^+{}_{mq}\; W_2^+{}_n{}^q.\label{eq:Ricci12}
\end{split}\end{equation}
The Ricci tensor as computed from the above torsion classes and directly from the metric agree and is given by:
In the coordinate basis,
\beqn
R_{mn} = 
\left(\begin{array}{cccccc}
						1&    0              &  0         & 0        & 0    & 0  \\
						0&1 -(x^3)^2 -(x^4)^2& -x^1 x^4   & x^1 x^3  & -x^3 & x^4\\
						0& -x^1 x^4          & 1 - (x^1)^2&       0  & 0    & x^1\\
						0&  x^1 x^3          & 0          &1 -(x^1)^2& x^1  & 0  \\
						0& -x^3              & 0          &x^1       &-1    & 0  \\
						0&  x^4              & x^1        & 0        & 0    &-1  \\
\end{array}\right),\label{I12Rtcb}
\eeqn
which in the tangent basis simplifies to
\beqn
R_{\bar{m}\bar{n}} =\,{\rm Diag}\{1,1,1,1,-1,-1\}. \label{I12Rt}
\eeqn
where barred indices refer to tangent space. Similarly, the torsion class computation of the Ricci scalar is consistent with the trace of the metric-produced Ricci tensor:
\beqn
R &=& R_{mn} g^{nm}\nolabel\\
  &=& \frac{15}{2} (W_1^+)^2- \frac{1}{4} W^+_2{}_{mn} W_2^+{}^{mn} \label{I12Rs}\\
  &=& 2 \label{I12Rsv}
\eeqn

The metric for the $\{W_3\}$ Iwasawa manifold of \cite{lopes} is produced by altering in (\ref{ds2w12}) the overall sign of the two terms in brackets with $dx^5$, and therefore simultaneously altering the signs of the terms on 
the right-hand sides of (\ref{I12e5}) and (\ref{I12de5}). The metric then becomes,
\beqn
ds^2 = \sum_{i=1}^4 (dx^i)^2 + (dx^5 + x^1 dx^4 - x^3 dx^2)^2 + (dx^6 - x^1 dx^3 - x^4 dx^2)^2,
\label{ds2w3}
\eeqn
with tangent basis
\beqn   
e^j &=& dx^j,\, j= 1,\, ...\, ,\, 4,\label{I3ej}\\
e^5 &=& dx^5 + x^1 dx^4 - x^3 dx^2, \label{I3e5}\\
e^6 &=& dx^6 - x^1 dx^3 - x^4 dx^2, \label{I3e6}
\eeqn 
and exterior derivatives
\beqn   
de^j &=& 0,\, j= 1,\, ...\, ,\, 4,\label{I3dej}\\
de^5 &=&  e^1 \wedge e^4 + e^2 \wedge e^3, \label{I3de5}\\
de^6 &=& -e^1 \wedge e^3 + e^2 \wedge e^4. \label{I3de6}
\eeqn

The only non-zero torsion class for this manifold is
\beqn
W_3 = \frac{1}{2}(e^1 + i e^2)\wedge (e^3 + i e^4)\wedge (e^5 - i e^6).
\label{I3W3}
\eeqn
Hence, the torsion expression for the Ricci tensor reduces to
\beqn
\begin{split}
R_{mn}(W_3)=& -\frac{1}{2}J_{(m}{}^p\nabla^s\left[W_3{}_{|sp|n)} + \overline{W}_3{}_{|sp|n)} \right]\\
            & +\frac{1}{4}\left[W_3{}_{muv} {\overline{W}}_3{}_n{}^{st} +W_3{}_{nuv}       
               {\overline{W}}_3{}_m{}^{st}\right] \Pi^+_s {}^u \Pi^+_t {}^v \\
            & - \frac{1}{2}\left[W_3{}^s{}_{mu} W_3{}^u {}_{pq} \Pi^+_n {}^p \Pi^+_s {}^q +\overline{W}_3{}^s{}_{mu} 
               \overline{W}_3{}^u {}_{pq} \Pi^-_n {}^p \Pi^-_s {}^q \right]\\
            & +\frac{1}{2} \left[ W_3{}_{suv} \Pi^+_{(m} {}^u \Pi^+_{|q|} {}^v \ol{W}_3^{sq}{}_{n)}  
               +\ol{W}_3{}_{suv} \Pi^-_{(m} {}^u \Pi^-_{|q|} {}^v W_3^{sq}{}_{n)}\right].  \label{eq:RicciI3}
\end{split}
\eeqn
The two computations of the Ricci tensor for this second Iwasawa manifold are again consistent. 
In the coordinate basis,
\beqn
R_{mn} = 
\left(\begin{array}{cccccc}
						1&    0              &  0         & 0        & 0    & 0  \\
						0&1 -(x^3)^2 -(x^4)^2& -x^1 x^4   & x^1 x^3  & x^4  & x^3\\
						0& -x^1 x^4          & 1 - (x^1)^2&       0  & x^1  & 0  \\
						0&  x^1 x^3          & 0          &1 -(x^1)^2& 0    &-x^1\\
						0&  x^4              & x^1        & 0        &-1    & 0  \\
						0&  x^3              & 0          & -x^1     & 0    &-1  \\
\end{array}\right),\label{I3Rtcb}
\eeqn
which in the tangent basis simplifies, as in (\ref{I12Rs}), to
\beqn
R_{\bar{m}\bar{n}} =\,{\rm Diag}\{1,1,1,1,-1,-1\}. \label{I3Rt}
\eeqn
The torsion class computation of the Ricci scalar is again consistent with the trace of the metric-produced Ricci tensor:
\beqn
R &=& R_{mn} g^{nm}\nolabel\\
  &=&  -\frac{1}{6} W_3{}_{pqr} \ol{W}_3{}^{pqr}\label{I3Rs}\\
  &=& 2 \label{I3Rsv}
\eeqn

The $\{W^+_1,\, W^+_2,\, W_3 \}$ two-step nilmanifold is generated by removing the $-x^1 dx^3$
in brackets with $dx^6$ in the metric (\ref{ds2w12}) and therefore removing the
$- x^1 dx^3$ term in (\ref{I12e6}) and the $-e^1 \wedge e^3$ term in (\ref{I12de6}). The metric becomes 
\beqn
ds^2 = \sum_{i=1}^4 (dx^i)^2 + (dx^5 - x^1 dx^4 + x^3 dx^2)^2 + (dx^6 - x^4 dx^2)^2.
\label{ds2w123}
\eeqn
with tangent basis defined by 
\beqn 
e^j &=& dx^j,\, j= 1,\, ...\, ,\, 4,\label{I123ej}\\
e^5 &=& dx^5 - x^1 dx^4 + x^3 dx^2, \label{I123e5}\\
e^6 &=& dx^6 \phantom{- x^1 dx^3} - x^4 dx^2, \label{I123e6}
\eeqn
and exterior derivatives
\beqn   
de^j &=& 0,\, j= 1,\, ...\, ,\, 4,\label{I123dej}\\
de^5 &=& -e^1 \wedge e^4 - e^2 \wedge e^3, \label{I123de5}\\
de^6 &=& \phantom{-e^1 \wedge e^3} + e^2 \wedge e^4, \label{I123de6}
\eeqn
This alteration (\ref{I12e6}) generates a manifold for which all three half-flat torsion classes are non-zero:
\beqn
W^+_1 &=& -\frac{1}{2} \label{I122W1}\\ 
W^+_2 &=& -\left[e^1 \wedge e^2 + e^3 \wedge e^4 - 2 e^5 \wedge e^6 \right], \label{I123W2}\\
W_3   &=& -\frac{i}{8}
           \left[3 e^1\wedge e^3\wedge e^5 + e^1\wedge e^4\wedge e^6 + e^2\wedge e^3\wedge e^6 + e^2\wedge e^4\wedge  
           e^5 \right]\nolabel\\
      &&  +\frac{1}{8}
           \left[3 e^2\wedge e^4\wedge e^6 + e^2\wedge e^3\wedge e^5 + e^1\wedge e^4\wedge e^5 + e^1\wedge e^3\wedge e^6 \right] \label{I123W3}. 
\eeqn
The torsion class and metric computations of the Ricci tensor are again matching for Ricci tensor and for Ricci
scalar. In the coordinate basis,
\beqn
R_{mn} = 
\left(\begin{array}{cccccc}
	\frac{1}{2}&    0              &  0         & 0        & 0    & 0  \\
						0&1 -(x^3)^2 -\frac{1}{2}(x^4)^2  & 0        & x^1 x^3  & -x^3 & \frac{1}{2}x^4\\
						0&    0              & \frac{1}{2}&  0       & 0    & 0  \\
						0&  x^1 x^3          & 0          &1 -(x^1)^2& x^1  & 0  \\
						0& -x^3              & 0          &x^1       &-1    & 0  \\
						0&  \frac{1}{2}x^4   & x^1        & 0        & 0    &-\frac{1}{2}x^4\\
\end{array}\right),\label{I123Rtcb}
\eeqn
or equivalently in the tangent basis,
\beqn
R_{\bar{m}\bar{n}} =\,{\rm Diag}\{\frac{1}{2},1,\frac{1}{2},1,-1,-\frac{1}{2}\}, \label{I123Rt}
\eeqn
and 
\beqn
R = \frac{3}{2}. \label{I123Rs}
\eeqn

\setcounter{equation}{0}
%
\section{Half-flat Manifolds from Mirror Symmetry}\label{sec:defCY}
In this section we review the results of \cite{gurrieri}, in which the authors found that a certain kind of half-flat manifolds arise naturally within the flux compactification schemes of string theory. These half-flat manifolds appear when mirror symmetry is extended to the realm of type II string theories with non-zero NS fluxes compactified on a background Calabi-Yau manifold, and thus they bear close connection to Calabi-Yau manifolds. We also present here, for the first time, explicit expressions for the Ricci tensor, as well as a \emph{new} constraint for the K\"ahler moduli of the compactification of string theory on these half-flat manifolds. 
\subsection{The Mirror duals of CY Manifolds with $NS$ Fluxes}
In \cite{gurrieri}  Gurrieri et al.\  made the conjecture that  the mirror symmetric dual of type IIB (A) string theory compactified on a Calabi-Yau $\W$ with `electric' NS  fluxes is given by IIA (B) string theory on a half-flat manifold $\tm$, which may be thought of as a kind of deformation of the Calabi-Yau manifold $\M$; this latter Calabi-Yau is the one which is the canonical mirror dual of $\W$ in the absence of fluxes.\footnote{We denote by $\tm$ \emph{only} those half-flat manifolds which may be thought of as deformations of a genuine Calabi-Yau manifold $\M$ as result of applying mirror symmetry to the Calabi-Yau $\W$ with NS fluxes. Note that because we have considerable choice in choosing the fluxes on $\W$, we should, in principle, denote the dual by $\tm_{\{e_i\}}$, where $\{e_i\}$ is the set of flux parameters, but since we do not deal with issues arising from this choice, we shall not overburden our notation and stick with $\tm$.} Gurrieri et al.\ also showed that the low energy effective action of IIA supergravity on $\tm$ is the same as IIB on $\W$ with NS fluxes, thus furnishing strong evidence for their conjecture.
\par
Although $\tm$ is half-flat, a complete characterization of this manifold and its exact correspondence to the `underlying' Calabi-Yau $\M$ is still not possible. However, at least in the large volume and `large complex structure' limit (the meaning of the latter limit is explained in more detail in the next sub-section), it still seems possible to compute most of the quantities relevant for $\mathcal{L}_{\mathrm{eff}}$ on this manifold by using the standard results on Calabi-Yau manifolds. 

In order to establish notation before we present our results regarding $\tm$, we shall review some of the results and methods of Gurrieri et al. Our aim is not to try to be exhaustive and the interested reader may look up \cite{gurrieri} for more details.
\par
For definiteness, consider type IIB on $\W$ with non-zero NS fluxes. These fluxes are characterized by a set of integers ${e_I}$  which on the type IIB side are given by
\begin{equation}
H^B_{3}=d\tilde{B}_{2}^B+  e_I \beta^I \label{eq:II}
\end{equation}
where $H^B_{3}$ is the ten-dimensional NS 3-form field strength and $\tilde{B}^B_2$ is the four-dimensional NS 2-form potential, and $\b^I$ with $I=0,1,\dots, h^{2,1}(\W)$, are half of the standard real symplectic basis of the third cohomology group $H^3(\W)$ of $\W$. \footnote{We shall not be too careful about distinguishing between real and integral cohomology.} (This, incidentally, is the meaning of `electric fluxes'; the coefficients of $\a_I$, the other half of the symplectic basis, would be the magnetic fluxes, but they are set to zero in \cite{gurrieri}. We also set the magnetic fluxes to zero. For the inclusion of magnetic fluxes see \cite{grana3}) It is also convenient to group the symplectic vectors $\b^I$ and their coefficients $e_I$ into two sets, the first consisting of $\b^0$ and $e_0$ and the second $\b^i$ and $e_i$ with $i=1,\dots, h^{2,1}(\W)$. We shall soon see the motivation for doing so, as they play quite different roles on the mirror manifold.
\par
Turning on fluxes on a Calabi-Yau background still preserves $\mathcal{N}=2$ supersymmetry of the low energy effective theory but now, instead of being a supergravity theory coupled to hyper-, tensor- and vector-multiplets, the theory becomes massive (i.e., a potential is induced), and some of the Peccei-Quinn symmetries become gauged. Because $\mathcal{N}=2$ supersymmetry of $\mathcal{L}_{\mathrm{eff}}$ depends solely on the structure group of the internal manifold being $SU(3)$, this singles out for the mirror candidate only those manifolds which have $SU(3)$ as their structure group. Calabi-Yau manifolds are of course a very special subset of this class for which the holonomy group of the Levi-Civita connection is also $SU(3)$. 
\par
In string theory on Calabi-Yau manifolds the K\"ahler form $J$ naturally combines with the internal components of the NS gauge potential $B$ to form a complex closed two form $B+iJ$. The moduli space relevant for string theory is the complexified K\"ahler cone and it is the parameter space of the deformation of this complexified two-form. However, when one turns-on NS fluxes, the real part of this two form is no longer closed. As mentioned above, the mirror manifold, because of $\mathcal{N}=2$ supersymmetry, must be an $SU(3)$ manifold and hence must have globally defined $\Om$ and $J$. However, $\Om$ and $J$ no longer have the interpretation of being the holomorphic three form that defines a Calabi-Yau uniquely, and the K\"ahler form, respectively; but now they are components of $SU(3)$ structure. In the familiar Calabi-Yau setting, mirror symmetry may be seen as exchanging these two $SU(3)$ structure forms between the two mirror-symmetric pair. Thus it seems natural to assume that in the more generalized case at hand mirror symmetry continues to do the same. Then on the presumed mirror of $\W$ with fluxes, the real part of $\Om$ can no longer be closed\footnote{There is some arbitrariness in choosing the real and imaginary part of $\Om$ but this is not central to our discussion at the moment.} as its mirror counterpart $B$ is no longer closed. Thus we see the mirror can no longer be a Calabi-Yau manifold. Let us denote this manifold by $\tm$ as we expect it to have some connection to the Calabi-Yau $\M$ which is the canonical mirror dual of $\W$.
\par
Since $SU(3)$ structure is classified by the torsion classes that appear in $dJ$ and $d\Om$, the authors of \cite{gurrieri} then proceeded to identify the various non-zero torsion classes of $\tm$ with the $SU(3)$ representations of the various cohomology groups of $\M$. The rationale for this is as follows. The various cohomology groups of a Calabi-Yau manifold furnish different representations of $SU(3)$. On the other hand, in the small flux limit $\tm$ should, at least locally, look like $\M$ and so all the forms which exist on the latter should also exist on the former furnishing the same representations of $SU(3)$. So it is sensible to identify these $SU(3)$ representations with corresponding torsion classes which appear in the exterior derivatives of $J$ and $\Om$ on $\tm$. This  heuristic argument was made more precise in \cite{gurrieri} in various steps and it was shown that the surviving torsion classes lead to $\tm$ being a half-flat manifold. Also it was shown that the standard basis of the cohomology groups of $\M$ should satisfy on $\tm$ the following relations:
\begin{equation}
d\a_0 = e_i \om^i,\;\;d\a_\a = d\b^a=0,\;\;d\om_i =e_i \b^0,\;\;d\tilde{\om}^i =0, \label{eq:rules}.
\end{equation}
As a consequence, due to the standard definition of $J$ and $\Om$ (see Appendix \ref{ap:cymod} for more details), we also have:
\begin{eqnarray}
dJ &=& v^i e_i \b^0 \label{eq:dj2} \\
d\Om^+ &=&e_i \tilde{\om}^i \label{eq:dom2}.
\end{eqnarray}
In the above equations we have $\a_a$ and $\b^a$, with $a=0,1,\dots, h^{2,1}(\M)$ and $\a,\b=1,\dots, h^{2,1}(\M)$, are the standard symplectic basis of the third cohomology group of the Calabi-Yau manifold $\M$. Similarly $\om_i$ and $\tilde{\om}^i$ are the basis for $H^{1,1}(\M)$ and $H^{2,2}(\M)$ respectively. $v^i$ are half of the coordinates on the K\"ahler moduli space of $\M$ and is defined by $J=v^i \om_i$. Note that in the above expressions $e_0$ is absent; it appears, after dualizing, as the space-time components of the NS field strength on type IIA theory and thus acts as a sort of `cosmological constant' in $\mathcal{L}_{\mathrm{eff}}$.
\par
Since we expect the kinetic terms in $\mathcal{L}_{\mathrm{eff}}$ to remain the same, it was further assumed in \cite{gurrieri} that the above-mentioned forms should continue to satisfy the same integral relations on $\tm$ as they do on $\M$. In other words we still have
\begin{equation}
\int_{\tm} \a_a \w \b^b = \d_a^b, \label{eq:int1}
\end{equation} 
\begin{equation}
\begin{split}
\int_{\tm} \a_a  \w \a_b  =\int_\M \b^a  \w \b^b=0 \label{eq:int2}
\end{split}
\end{equation}
and 
\begin{equation}
\int_{\tm} \om_i \w \tilde{\om}^j = \d_i{}^j. \label{eq:int3}
\end{equation}
However, since these forms are no longer harmonic due to (\ref{eq:rules}), they do not have the same topological significance for $\tm$ as they did for $\M$. Indeed it was noted in \cite{gurrieri} that the dimensions of the second and the third cohomology group of $\tm$ changes with respect to those of $\M$, but in a way that the Euler characteristics of the two manifolds are the same. Note that the expressions in (\ref{eq:rules}) -- (\ref{eq:dom2})  are given in a fixed special geometric gauge.
\par
The above rules for taking exterior derivatives and performing integrals on $\tm$, are only valid in the large volume limit (i.e.\ the supergravity limit) \emph{and} the large complex structure limit. The latter condition comes from the fact that on the IIB side one is in the large volume limit, since the gravitational back-reaction on the geometry due to the fluxes is ignored, and hence on the IIA side, by mirror symmetry, one must be in the large complex structure limit. Using the above rules the authors of \cite{gurrieri} then derived the effective action of type IIA on $\tm$ and showed that it is the same as the effective action of IIB  compactified on the Calabi-Yau $\W$ with NS fluxes.
\par
Note that the dimensional reduction in \cite{gurrieri} could have been carried out, in principle, using a  harmonic basis on $\tm$ but it is not clear that that basis represents the physically meaningful degrees of freedom. Thus from a physical point of view it is more convenient to continue using the language of Calabi-Yau compactifications and interpret the extra terms that arise from the non-closure of the forms as being  terms for the effective potential generated for the moduli as a result of fluxing\footnote{By `fluxing' we mean `to introduce fluxes into a geometry erstwhile without fluxes'.}. This is also the natural language in which to make connection with gauged supergravity. Thus we continue to call the  scalar fields, familiar from CY compactification, `moduli' but now they come with a non-vanishing potential, which fixes them partially.  
\subsection{Torsion Classes and the Ricci Curvature}
Let us now deal with the issue of `large complex structure' alluded to before. What is the meaning of the large complex structure limit for a manifold which is not complex? Recall that at a generic point on the complex structure moduli space of a Calabi-Yau, the pre-potential is a complicated, homogeneous holomorphic function of degree two. But in the large complex structure limit this function becomes a simple polynomial. Thus, we interpret the above statement by assuming that the prepotential of the complex structure moduli space takes this latter simple form:
\begin{equation}
\mathscr{G}= -\frac{1}{3!} \frac{\kappa_{\a\b\g}z^\a z^\b z^\g}{z^0} + \frac{1}{2} S_{\a\b}z^\a z^\b + z^0 A_\a z^\a. 
\end{equation}
where $\{z^0, z^\a\}$ are projective coordinates and following \cite{gurrieri} we set $z^0=1$ in what follows.
\par
We now derive expressions for the torsion classes from eqs. (\ref{eq:dj2}) and (\ref{eq:dom2}) in  the language of the moduli space of Calabi-Yau manifolds \cite{candelas,bodner}. To do so we recall the expression for $\b^0$ in the large complex structure limit (see Appendix \ref{ap:cmplxlarge} for more details):
\begin{equation}
\beta^0 = \frac{1}{4\V} \Om^- + \frac{y^\a}{4\V} \left(\chi_\a +\bar{\chi}_{\bar{\a}}\right)
\end{equation}
where $y^\a$ is the imaginary part of the complex structure modulus $z^\a$, and $\chi_\a$ and $\bar{\chi}_{\bar{\a}}$ span $H^{2,1}$ and $ H^{1,2}$ of $\M$, respectively. Using the above identification it is easy to show that
\begin{equation}\begin{split}
W_1^+ &= \frac{1}{6} \frac{e\cdot v}{\V} \\
W_2^+ &= - \frac{e\cdot \om}{4\V} +\frac{1}{3} \frac{e\cdot v}{\V} J\\
W_3 &= \frac{e\cdot v}{4\V} \chi  
\end{split}\end{equation}
where we have introduced the notation
\begin{equation}\begin{split}
y^\a \chi_\a &\equiv \chi\\
e\cdot v &\equiv e_i v^i\;\;\mathrm{etc.}
\end{split}
\end{equation}
Now that we have identified the individual torsion classes we can use eq.(\ref{eq:Ricci}) to find expressions for the Ricci tensor and the Ricci scalar of $\tm$. We choose to express our results in complex coordinates and thus we have
\begin{eqnarray}
R_{\mu \bar{\nu}}&=& \frac{e^2}{16\V^2}g_{\mu\bar{\nu}} -\frac{3}{16} \frac{(e\cdot v)^2}{\V^2} g_{\mu\bar{\nu}} 
- \frac{3i}{16} \frac{e\cdot v}{\V^2} (e\cdot \om)_{\mu\bar{\nu}} - \frac{1}{32\V^2} (e\cdot \om)_{\mu\bar{\rho}} (e\cdot \om)_{\bar{\nu}}{}^{\bar{\rho}}\nonumber \\
&&+ \frac{(e\cdot v)^2}{64\V^2} \bar{\chi}_{\mu\bar{\sigma}\bar{\tau}} \chi_{\bar{\nu}}{}^{\bar{\sigma}\bar{\tau}} - \frac{(e\cdot v)^2}{32\V^2} \chi_{\mu\theta \bar{\tau}}\bar{\chi}_{\bar{\nu}}{}^{\theta\bar{\tau}} \label{eq:mixed} \\
&&\nonumber \\
R_{\mu\nu}&=& -i\frac{e^2}{128\V^2} (\bar{\chi}\cdot \Om)_{\mu\nu} + \frac{(e\cdot v)^2}{32\V^2} \chi_{\mu}{}^{\psi\bar{\sigma}}\chi_{\nu\psi\bar{\sigma}}\nonumber \\
&&+ \frac{(e\cdot v)}{128\V^2}\left[\frac{1}{2} (e\cdot\om)_{\mu}{}^{\rho} (\bar{\chi}\cdot \Om)_{\nu\rho} + (e\cdot\om)_{\theta}{}^{\rho} \bar{\chi}_{\mu}{}^{\psi\theta} \Om_{\nu\psi\rho} \right]\label{eq:pure}
\end{eqnarray}
and the Ricci scalar is
\begin{equation}\begin{split}
R=& \frac{3e^2}{8\V^2}+\frac{3(e\cdot v)^2}{8\V^2} -\frac{1}{16\V^2} (e\cdot \om)_{\mu\bar{\rho}} (e\cdot \om)^{\mu\bar{\rho}} - \frac{1}{32} \frac{(e\cdot v)^2}{\V^2} \chi_{\mu\theta\bar{\tau}} \bar{\chi}^{\mu\theta\bar{\tau}} \label{eq:ricciscalar}
\end{split}\end{equation}
and so we have
\begin{equation}
\int_{\tm} d^6y\sqrt{g}\;R  = \frac{1}{8} \frac{e^2}{\V} \label{eq:intricci}
\end{equation}
which is in agreement with the result derived in \cite{gurrieri} when one takes into account the gauge choice $\|\Om\|^2= 8$.
\par
Next, we shall use a consistency argument involving eq.\ (\ref{eq:intricci}) to establish whether we can derive any new condition on $\tm$ that is not part of the rules, (\ref{eq:rules}), proposes in \cite{gurrieri}. To do this we shall consider the variation of (\ref{eq:intricci}). The left hand side of this variation will involve the explicit expressions, (\ref{eq:mixed}) and (\ref{eq:pure}), for the Ricci tensor and we can evaluate these integrals using properties of CY moduli spaces, as was done in deriving the effective action in \cite{gurrieri}. On the other hand the variation of the right hand side of (\ref{eq:intricci}) may also be evaluated by using the standard results of CY moduli spaces. These two expressions, of course, should be identical. If not, then any new rules that we may derive in imposing this equivalence, should then be appended to the list (\ref{eq:rules}). 

On a manifold in which we compute the Ricci curvature from the metric, this identity is automatically satisfied. But for the case at hand, we only have indirect information about the geometric quantities on $\tm$ and it may be that the rules (\ref{eq:rules}) are not exhaustive to ensure that such identities are automatically satisfied. Thus such consistency arguments are valuable since they furnish us with further information regarding these manifolds. Indeed, below we find a new constraint on the K\"ahler moduli space in this way. Note that there may be other such consistency checks that may yield new, further constraints on the moduli spaces.
For the sake of greater readability, details of the derivations of many of the formulae used in the following pages are relegated to Appendix \ref{ap:details}. 

Returning to the variation of (\ref{eq:intricci}) we get
\begin{equation}\begin{split}
\int_{\tm} \sqrt{g} d^6y \;\delta g^{mn}\; R_{mn}= \frac{1}{8\V} \d e^2- \frac{e^2}{\V} (v\cdot \d v) \label{eq:varricci}
\end{split}
\end{equation}
where the term with $\d R_{mn}$ drops out because it is a total derivative.\footnote{Checking that the $\d R_{mn}$ term is a total derivative would be another consistency check on the rules (\ref{eq:rules}) -- (\ref{eq:int3}), but we leave that as a future exercise.} In the above equation we have $\d e^2 = e_i e_j \d G^{ij}$, where $G^{ij}$ is the inverse of the metric on the K\"ahler moduli space, and $(v\cdot \d v) = v^i \d v^j G_{ij}$. More details regarding the derivation of (\ref{eq:varricci}) can be found in Appendix \ref{ap:details}. Note that, the above expression is given in special coordinates and is \emph{not} coordinate invariant. 
\par
Let us choose our moduli fields to be related to the variation of the metric by
\begin{equation}\begin{split}
\delta g_{\mu\bar{\nu}} &= -i \d v^i (\om_i)_{\mu\bar{\nu}}\\
\delta g_{\mu\nu} &=  \frac{1}{8} (\Om\cdot {\bar{\chi}}_{\bar{\alpha}})_{\mu\nu} \delta {\bar{z}}^{\bar{\alpha}}. \label{eq:moduli}
\end{split}\end{equation}
In deriving the effective action there is an arbitrariness in choosing the phase in the second of these equations \cite{candelas,bodner}. Our choice here is motivated by two facts. First, the right hand side of (\ref{eq:varricci}) is entirely in terms of variation of K\"ahler moduli, and hence the terms on the left hand side that involve variation of the complex structure moduli must be transformed to the appropriate variation of the K\"ahler moduli. For the conventions that we are using ($z^0=1$ and $\|\Om\|^2=8$) it can be shown that the prefactor above has to be $\pm 1$. Secondly, the condition that we get below (eq.(\ref{eq:infcond})) from (\ref{eq:varricci}) this way, we apply to the case where the underlying Calabi-Yau $\M$ has $h^{1,1}(\M)=1$ (e.g., the famous quintic has this property) and in this case, as shall be shown below, the dependence of (\ref{eq:varricci}) on the flux parameters $\{e_i\}$ drops out. The resulting equation in this limiting case must be an identity on the moduli space of the Calabi-Yau $\M$ with  $h^{1,1}(\M)=1$. This further fixes the phase to be $+1$. We shall return to this second point below.
\par
Using (\ref{eq:moduli}) and (\ref{eq:mixed}) we find for mixed part of the l.h.s.\ of (\ref{eq:varricci})
\begin{equation}\begin{split}
2\int_{\tm} \sqrt{g} d^6 y \; \delta g^{\mu \bar{\nu}} R_{\mu \bar{\nu}}&= -\frac{1}{2} \frac{(e\cdot v)^2}{\V} (v\cdot \d v) -\frac{1}{2} \frac{(e\cdot v)}{\V} (e\cdot \d v) + \frac{\K_{ijk}}{32 \V^2} \d v^i e^j e^k\\
&+\frac{1}{16} \frac{(e\cdot v)^2}{\V^2} \int_{\tm} d^6\xi \delta v^i \left[\frac{1}{2} (\om_i)^{\mu{\bar{\nu}}} \bar{\chi}_{\mu \bar{\sigma} \bar{\tau}} \chi_{\bar{\nu}}{}^{\bar{\sigma} \bar{\tau}} - (\om_i)^{\mu\bar{\nu}} \chi_{\mu \sigma \bar{\tau}} \bar{\chi}_{\bar{\nu}}{}^{\sigma \bar{\tau}} \right]\\ 
&\equiv X
\end{split} \label{eq:intX}
\end{equation}
where $e^i = G^{ij} e_j$, and $\K_{ijk}$ are the Yukawa couplings in the K\"ahler sector (defined in Appendix \ref{ap:cymod}) and we have introduced complex coordinates $\xi^\mu$ and $\bar{\xi}^{\bar{\mu}}$ such that $d^6y = i d^6\xi\equiv i d^3\xi d^3\bar{\xi} $. By noting that all six forms are proportional it is easy to evaluate the integrals on the second line above and thus we find 
\begin{equation}\begin{split}
X &=  -\frac{1}{2} \frac{(e\cdot v)}{\V} (e\cdot \d v) + \frac{1}{32 \V^2} \K_{ijk} \d v^i e^j e^k. \label{eq:Xxpression}
\end{split}\end{equation}
Similarly, by using (\ref{eq:moduli}) and (\ref{eq:pure}), we find for the pure part of the l.h.s.\ of (\ref{eq:varricci})
\begin{equation}\begin{split}
\int_{\tm} \sqrt{g} d^6 y \; \left( \d g^{\mu\nu} R_{\mu\nu}+ \d g^{\bar{\mu}\bar{\nu}} R_{{\bar{\mu}} {\bar{\nu}}}\right)& =-\frac{e^2}{2\V}  \d y^\a y^\b G_{\a\bar{\b}}       \\
&\equiv Y
\label{eq:cmplxvar}
\end{split}
\end{equation}
where $G_{\a\bar{\b}}$ is the hermitian metric on the complex structure moduli space of $\M$. More details on evaluating these integrals can be found in Appendix \ref{ap:integrals}.
\par
As mentioned above, the expression for $X+Y$ that we get this way involves moduli from both the K\"ahler and the complex structure sectors. To compare to the r.h.s. of (\ref{eq:varricci}) we must be able to transform the expression (\ref{eq:cmplxvar}) into a statement about the variation on the K\"ahler moduli space. To do this note that as a consequence of $SU(3)$ structure we have
\begin{equation}
\Om \w \overline{\Om} = -\frac{4i}{3} J\w J \w J.
\end{equation}
The left hand side of this relation only depends on complex structure while the right hand side on K\"ahler structure. Varying both sides and using Kodaira's formula (\ref{eq:kodaira}), and then integrating we get, in the gauge we are using,
\begin{equation}
G_{\a\bar{\b}} \d y^\a  y^\b= G_{ij} v^i \d v^j.
\end{equation}
This way we get
\begin{equation}\begin{split}
X+Y &= -\frac{3}{2} \frac{(e\cdot v)}{\V} (e\cdot \d v) + \frac{\d e^2}{8\V} + 2 \frac{(e\cdot v)^2}{\V} (v \cdot \d v) - \frac{e^2}{\V} (v\cdot \d v)
\end{split}\end{equation}
where we have used (\ref{eq:anotherident}) to transform the $\mathcal{K}_{ijk} e^i e^j \d v^k$ term in (\ref{eq:Xxpression}). Putting all of this together in (\ref{eq:varricci}) we finally find the advertised constraint on the K\"ahler moduli space:
\begin{equation}
(e\cdot v)^2 (v\cdot \d v) = \frac{3}{4} (e\cdot v) (e\cdot \d v). \label{eq:infcond}
\end{equation}
Now we choose $\M$ to be a Calabi-Yau manifold with $h^{1,1}(\M)=1$. Such $\M$ are allowed since the mirror $\W$ then has $h^{2,1}=1$ and two flux parameters $e_0$ and $e_1$ enter in the picture. However, as mentioned above, of these only the latter enter into $e\cdot v$ and thus drop out of (\ref{eq:infcond}).  Then using $G_{vv}= \frac{3}{4} v^{-2}$, the above equation becomes a trivial identity. This is  a consistency check for eq.(\ref{eq:infcond}).

Next by using the identity (which is derived in Appendix \ref{ap:details})
\begin{equation}\begin{split}
\d \frac{1}{\V}= - \frac{4}{\V} (v\cdot \d v)
\end{split}\end{equation}
we find that eq.(\ref{eq:infcond}) implies
\begin{equation}\begin{split}
e\cdot v = \mathcal{C} \V^{\frac{1}{3}} \label{eq:intcond}
\end{split}\end{equation}
where $\mathcal{C}$ is a constant on the moduli space. Note that eqs.(\ref{eq:infcond}) or (\ref{eq:intcond}) are valid for any $\tm$, not just for the ones corresponding to $h^{1,1}(\M)=1$.
\par
From this expression it is easy to find out how the volume of the internal space changes as a function of the `seventh coordinate' $z$ (not to be confused with the complex structure moduli) of the $G_2$ generalized cylinder. Using Hitchin's flow equation (\ref{eq:flow2}) we find
\begin{equation}\begin{split}
\V= \V_0 \left( 1-\frac{z}{z_0}\right)^\frac{3}{2} \label{eq:volflow}
\end{split}\end{equation}
where $z_0$ is another integration constant and 
\begin{equation}
\V_0 = \left(\frac{\mathcal{C}z_0}{3}\right)^\frac{3}{2}. 
\end{equation}
This implies that the particular supergravity approximation we have made is only valid in the region away from $z\approx z_0$.
\setcounter{equation}{0}
%
%
\section{Discussions and Conclusions}
This paper grew out of an effort to gain a deeper understanding of the role of the half-flat manifolds $\tm$ in string theory and their relationship to the Calabi-Yau manifold $\M$. Ideally one should be able to understand the theory that arises from $\tm$ \emph{independently} of $\M$ as well as having a clear idea of how they are related. So far we know how to compute $\mathcal{L}_{\mathrm{eff}}$ on $\tm$ only by extending our intuitions of $\M$ within a very limited regime. Even in that regime it is not clear to us that all that can be computed on $\tm$ has been computed. Our computation of the Ricci curvature of $\tm$ is another step toward obtaining the complete list of objects that we can compute, after $\mathcal{L}_{\mathrm{eff}}$ \cite{gurrieri}. The computation of the Ricci curvature of $\tm$ has not been discussed in the literature before. The general expression that we derived for the Ricci curvature of a generic half-flat manifold (not just $\tm$) has been independently computed in \cite{bedulli}. By using the expression for the Ricci curvature of $\tm$ we have found, by imposing certain consistency requirements,  a very simple and natural constraint, (\ref{eq:infcond}) or (\ref{eq:intcond}), on the `K\"ahler' moduli space. We believe this constraint has not been discussed in the literature before. Hitchin's flow equations then imply how the volume (and hence the effective four dimensional Newton's constant) varies with the direction transverse to a domain wall (which is the BPS solution of $\mathcal{L}_{\mathrm{eff}}$). This formula, (\ref{eq:volflow}), for the volume seems to crop-up whenever one tries to incorporate fluxes in a Calabi-Yau manifold. It had made its appearance in M-theory \cite{ali} and heterotic M-theory \cite{krause} before.
\par
It is well known by now that when fluxes are introduced (whether they originate from the `matter' sector \cite{louis,dallagata} or the `geometric' or `NS' sector \cite{gurrieri}) $\mathcal{L}_{\mathrm{eff}}$ is transformed from supergravity coupled to matter multiplets to a \emph{gauged, massive} supergravity. The effect of the induced potential is to fix some of the moduli.  Here we have found a certain relation that eliminates one of the moduli from the K\"ahler moduli space. However, this relationship becomes trivial when the underlying Calabi-Yau has just one modulus in the K\"ahler sector. Thus this can not be used to fix the volume modulus of the compactification. It would be interesting to see how this condition is related to moduli-fixing via the generated potential.
\par
In \cite{gurrieri} (also see \cite{gurrieri2}) a Gukov-Vafa-Witten type of superpotential \cite{gukov} was proposed, which plays an important role in the non-perturbative aspects of $\mathcal{N}=2$ compactifications of M-theory and string theory. This potential has the form \cite{gurrieri}
\begin{equation}
W| \propto \int_{\tm} \left(B+iJ\right)\w d\Om
\end{equation}
and it was evaluated, in the limit considered here, to be
\begin{equation}
W|\propto e_i s^i
\end{equation} 
where $W|$ denotes the projection of $W$ (which is defined on superspace) to the lowest spin component and $s^i = u^i + i v^i$ with $u^i$ being the moduli corresponding to NS potential $B= u^i \om_i$ (see Appendix \ref{ap:cymod} for more details). Due to our calculation we are able to present a different form of this result. It is easy to see using (\ref{eq:intcond}) that we have
\begin{equation}
W|\propto e_i u^i + i\mathcal{C} \V^{\frac{1}{3}}.
\end{equation}
Recently, an old non-renormalization theorem of string theory \cite{dine} was generalized \cite{burgess} to take into account the non-perturbative aspects of the  flux-com\-pact\-i\-fi\-ca\-tion of \cite{giddings} which has played an important role in the landscape scenario. It would be an interesting exercise to see whether some of the methods developed in \cite{burgess} can be applied to the case at hand.
\par
Next, let us discuss some of the subtleties left open by the works on this subject so far and what ought to be done to fill in the gaps. Following the lead of \cite{gurrieri} we have used, in Section \ref{sec:defCY}, many formulae whose validity is strictly true only on Calabi-Yau manifolds. These formulae don't seem to be in contradiction with the new modifications suggested by (\ref{eq:rules}) and (\ref{eq:intcond}). However, these formulae (of which Strominger's or Kodaira's formulae are just two) have deep geometric and topological significance for Calabi-Yau manifolds. Thus it seems to us that it is important to find how these formulae should be interpreted for the half-flat manifold $\tm$.

An important step in this direction was taken by a recent paper \cite{minasian} in which the authors try to make explicit the minimal set of geometric assumptions that are inherent in deriving $\mathcal{L}_{\mathrm{eff}}$ from compactification on a manifold with $SU(3)$ structure. Other more formal developments have been made in \cite{grana2,grana3}. Very briefly, these works explore a generalized set-up, based on the idea of `pure spinors', that is natural for mirror symmetry with fluxes. These works parallel the introduction, by Hitchin \cite{hitchin2}, of `generalized Calabi-Yau' manifolds in mathematics. On a  generalized Calabi-Yau manifold the almost complex and symplectic structures are considered to be a unitary object. Complex, symplectic and Calabi-Yau manifolds may be thought of as special cases when the various structures become integrable. (More details of this construction can be found in \cite{gualtieri}.)

Although, we hope to return to some of the issues left open by the present work in the light of some of the approaches mentioned above, we suggest here another approach, which might also lead to deeper insights into the structure of $\tm$ (as opposed to the more general $SU(3)$ manifolds).  A while ago Vafa and Shatashvili had initiated a study into the CFT of strings on manifolds with $G_2$ (and other exceptional) holonomy \cite{shatashvili}. The half-flat manifolds $\tm$ do not occur as a solution of string theory but rather the natural solution is of the form $\mathbb{R}^3\times \Z$ where $\Z$ is a generalized cylinder with $G_2$ holonomy. Thus to go beyond the various approximations that encumber one's progress, it might be a good idea to revisit some of the results obtained in \cite{shatashvili} for the generalized cylinder $\Z$ and try to rephrase them, with applications to half-flat manifolds in mind. This is perhaps quite an ambitious project but the gains might be worth the efforts.

One of the aspects of string theory today is that a string vacuum (in which we also include members of the `flux vacua') can not be understood in isolation. An isolated solution of string theory doesn't seem to make sense.  It is important to understand these new solutions, not only \emph{ab initio} but also \emph{in relation} to other `nearby' solutions. It seems to us that an approach solely based on $\mathcal{L}_{\mathrm{eff}}$ can only go so far, and that a more CFT approach is needed to understand this space of solutions. Of course for many of the solutions that are being considered today, the CFT approach is not enough since these solutions involve many non-perturbative effects.
\par
We would like to end this paper with a remark on the compactification of $E_8 \times E_8$ heterotic string theory on $\tm$. Since $\tm$ are closely related to $\M$, it is natural to enquire how these new manifolds are related to the famous phenomenology programme of string theory that began in \cite{candelas2}. Some important steps have already been taken in that direction \cite{gurrieri3, carlos} but there is much work to be done. Important for possible heterotic applications is that, due to the Green-Schwarz anomaly, stringy aspects (i.e., $\a'$ corrections) are expected to play an important role \cite{candelas2,strominger3}. So far there has been very little investigation of those aspects of $\tm$ that are relevant beyond the specific supergravity approximation that we also adopted here. This, again, brings us back to the $G_2$ picture alluded to above (which is also central to the derivations of the present paper). We believe some of the methods used in this paper can be extended to include some of those stringy aspects. This question is currently under study by the present authors \cite{ali2}.
\setcounter{equation}{0}
%
%
\section*{Acknowledgements}
It is a pleasure to thank Andrei Micu for many helpful discussions and communications regarding Section \ref{sec:defCY} of this paper. We would also like to thank him for bringing reference \cite{bedulli} to our attention. We would also like to thank Sigbj\o rn Hervik for his comments on the first version that helped us to correct a typo in equation (\ref{rctdef}).

\appendix
\renewcommand{\theequation}{\Alph{section}.\arabic{equation}}
\section{Differential Forms} \label{ap:difforms}
In this appendix we collect our conventions for forms. A $p$ form $A_p$ is defined to be
\begin{equation}
A_p= \frac{1}{p!}A_{m_1\dots m_p} dx^{m_1}\w\dots\w dx^{m_p}. \label{eq:norm}
\end{equation}
The Hodge star operator $\ast$ is defined by it action on the basis of $p$ forms
\begin{equation}
\ast\left( dx^{m_1}\w\dots \w dx^{m_p}\right)=\frac{\sqrt{|g|}}{(n-p)!}\;\e^{m_1\dots m_p}{}_{m_{p+1}\dots m_n}dx^{m_{p+1}}\w \dots\w dx^{m_n} 
\end{equation}
where $n$ is the total number of dimensions of spacetime, $|g|$ is the modulus of the determinant of the metric tensor and $\e_{m_1\dots m_n}$ is the completely antisymmetric tensor density in $n$ dimensions. Thus the action of performing Hodge duality twice has the following effect
\begin{equation}
\ast^2 A_p= (-1)^{p(n-p)+t} A_p
\end{equation}
where $t$ is 1 if the manifold is Lorentzian and 0 if it is Riemannian.
\par
If we have two $p$ forms $A_p$ and $B_p$ then we have
\begin{equation}
A_p\w \ast B_p = \frac{\sqrt{|g|}}{p!} A_{m_1\dots m_p} B^{m_1\dots m_p} dx^1\w\dots\w dx^n.
\end{equation}
If we set $C_{p+q}=A_p\w B_q$ then as a consequence of the normalization (\ref{eq:norm}) we have
\begin{equation}
C_{m_1\dots m_{p+q}}=\frac{(p+q)!}{p!q!} A_{[m_1\dots m_p} B_{m_{p+1}\dots {p+q}]}.
\end{equation}
\par
Now let $F_{p+1}=dA_p$ be a $p+1$ form, where $d$ is the exterior derivative. It is defined to be
\begin{equation}
dA_p=\frac{1}{p!}\partial_q A_{m_1\dots m_p} dx^q\w dx^{m_1}\w\dots\w dx^{m_p}
\end{equation}
where $\partial_q\equiv \frac{\partial}{\partial x^q}$. Thus we have
\begin{equation}
F_{q m_1\dots m_{p}}=(p+1) \partial_{[q}A_{m_1\dots m_p]}.
\end{equation}
%
%
\section{Calabi-Yau Moduli Spaces}\label{ap:cymod}
In this appendix we collect some facts about the  moduli spaces of Calabi-Yau three-folds. We do not aim to present a complete description of these spaces but rather to list the mathematical results used in the main text (in the context of half-flat manifolds) and also to set-out our conventions. Our main sources for this appendix are \cite{candelas, bodner, strominger, strominger2, ferrara}.
\par
The moduli of a Calabi-Yau three-fold $\M$ naturally separates into two classes: K\"ahler class deformations and complex structure deformations. We describe relevant aspects of the spaces of these two types of deformations below.
\subsection{K\"ahler Deformations}\label{ap:kahler}
The deformations of the K\"ahler structure of a Calabi-Yau threefold $\M$ are in one-to-one correspondence with the elements of $H^{1,1}(\M,\mathbb{Z})$ and are given in terms of the harmonic basis of that cohomology group. However, in string theory it is natural to complexify the K\"ahler form by adding to it the components of the NS two-form $B$ with its indices tangent to $\M$
\begin{equation}
B+iJ= \om_i s^i \;\;\mathrm{with}\;\;s^i = u^i + i v^i
\end{equation}
where $\om_i$, with $i=1,\dots, h^{1,1}(\M)$, are the basis of $H^{1,1}(\M,\mathbb{Z})$ with $h^{1,1}(\M)$ the corresponding Hodge number. $s^i$ are complex parameters which in string theory become dynamical fields in the four dimensional effective action.
\par
The space $H^{2,2}(\M,\mathbb{Z})$ is of course isomorphic to $H^{1,1}(\M,\mathbb{Z})$ and we can choose a dual basis $\{\tilde{\om}^i\}$ for the former in the following sense
\begin{equation}
\int_{\M} \om_i \w \tilde{\om}^j = \d_i{}^j. \label{eq:dualbasis}
\end{equation} 
\par
Some frequently used quantities, in the description of the effective actions for both type II and heterotic string theory compactifications, are
\begin{equation}
\begin{split}
\mathcal{K} &= \int_{\M} J\w J\w J \\
\mathcal{K}_i &= \int_{\M} \om_i\w J\w J \\
\mathcal{K}_{ij} &= \int_{\M} \om_i\w \om_j\w J \\
\mathcal{K}_{ijk} &= \int_{\M} \om_i\w \om_j\w \om_k.
\end{split}
\end{equation}
The volume of the Calabi-Yau is then given by
\begin{equation}
\mathscr{V}=\frac{1}{3!}\mathcal{K}. \label{eq:voldef}
\end{equation}
The metric that is naturally induced on the space of two forms is given by 
\begin{equation}
G_{ij} =\frac{1}{4\mathscr{V}}\int_{\M} \om_i \w * \om_j \label{eq:kmetric}
\end{equation}
and it can be shown using the complex coordinates $s^i$ introduced above that this metric is a K\"ahler metric. It is straightforward to derive 
\begin{equation}
*\om_i = - J\w \om_i  +\frac{3}{2} \frac{\K_i }{\mathcal{K}} J\w J.
\end{equation}
This may be used to express the metric as
\begin{equation}
G_{ij} = \frac{1}{4\mathscr{V}}\left[ -\mathcal{K}_{ij} + \frac{3}{2} \frac{\mathcal{K}_i \mathcal{K}_j}{\mathcal{K}}\right]. \label{eq:modkahlermetric}
\end{equation}
The inverse metric is given by
\begin{equation}
G^{ij} = -4\V \left[ (\K^{-1})^{ij} - \frac{1}{2\V} v^i v^j\right]. \label{eq:invkahler}
\end{equation}
Where $(\K^{-1})^{ij}$ is the inverse of the symmetric matrix $\K_{ij}$.
From (\ref{eq:dualbasis}) and (\ref{eq:kmetric}), the bases of $H^{1,1}(\M)$ and $H^{2,2}(\M)$ are related by
\begin{equation}
*\om_i = 4 \V G_{ij}\tilde{\om}^j.
\end{equation}
A useful formula involving the K\"ahler moduli that was first derived by Strominger in \cite{strominger} is
\begin{equation}
J_{mn} (\om^i)^{mn} = 8 v^i \label{eq:strominger}
\end{equation}
which is sometimes also expressed as
\begin{equation}
\frac{\mathcal{K}_i}{\V} = 8 v_i. \label{eq:strominger2}
\end{equation}
%
\subsection{Complex Structure Deformations}\label{ap:cmplx}
The complex structure deformations of $\M$ are in one-to-one correspondence with the elements of $H^{2,1}(\M,\mathbb{Z})$. A Calabi-Yau three-fold is defined by a holomorphic closed three-form $\Om$ (or its complex conjugate $\bar{\Om}$) which spans
$H^{3,0}(\M,\mathbb{Z})$ ($H^{0,3}(\M,\mathbb{Z})$) which is necessarily one dimensional.  $\Om$ is only defined up to a holomorphic function of the moduli $z^{a}$ with $a=0,1,\dots, h^{2,1}(\M,\mathbb{Z})$ which have been taken to be projective coordinates. In fact on the moduli space $\Om$ is not a scalar but a section of a $GL(1,\mathbb{C})$ line-bundle which is Hodge (i.e., its K\"ahler form is equal to the first Chern class.)
\par
On the space of three-forms one can introduce a hermitian inner-product 
\begin{equation}
\langle \a | \b \rangle = i \int_{\M} \a \w \b
\end{equation}
for $\a,\b\in H^{3}(\M)$ which enables one to introduce the following real basis for $H^3(\M,\mathbb{Z})$ consisting of $\a_a, a=0,\dots h^{2,1}(\M)$ and $\b^a, a=0,\dots h^{2,1}(\M)$ which satisfy
\begin{equation}
\int_{\M} \a_a \w \b^b = \d_a^b
\end{equation} 
and 
\begin{equation}
\begin{split}
\int_\M \a_a  \w \a_b  =\int_\M \b^a  \w \b^b=0  .
\end{split}
\end{equation}
These relations are $Sp(2h^{2,1}+2)$-invariant and hence the moduli space is in fact an $Sp(2h^{2,1}+2)\otimes GL(1,\mathbb{Z})$ bundle. In fact it can be shown that all quantities of interest on this manifold can be derived from a certain holomorphic function $\mathscr{G}(z)$ called the \emph{prepotential}. Such manifolds are known to be Special K\"ahler Manifolds. On such manifolds there is a choice of coordinates called special coordinates in which $\Om$ take the following form
\begin{equation}
\Om = z^a \a_a - \mathscr{G}_a(z)  \b^a
\end{equation}
where $\mathscr{G}_a(z) =\frac{\partial \mathscr{G}}{\partial z^a}$. A basis for $H^{2,1}$ is then given by Kodaira's formula,
\begin{equation}
\chi_a = \mathscr{D}_a \Om, \label{eq:kodaira}
\end{equation}
where we have introduced a $GL(1,\mathbb{C})$-covariant derivative by
\begin{equation}
\begin{split}
\mathscr{D}_a &= \frac{\partial}{\partial z^a} - k_a \\
k_a &= \frac{\int_{\M} \frac{\partial \Om}{\partial z^a }\w \bar{\Om}}{\int_{\M} \Om\w \bar{\Om}}.
\end{split}
\end{equation}
The gauge potentials $k_a$ can be shown to be derived from the K\"ahler potential $\mathbb{K}$ via
\begin{equation}
k_a =-\frac{\partial \mathbb{K}}{\partial z^a}
\end{equation} 
with 
\begin{equation}
e^{-\mathbb{K}}= i \int_\M \Om \w \bar{\Om}.
\end{equation}
One can thus identify the $GL(1,\mathbb{C})$ transformations as K\"ahler transformations. $\mathbb{K}$ defines a natural K\"ahler metric on the moduli space which is given below. Note that, only $h^{2,1}$ of the set, $\{\Phi_a, a = 0,\dots, h^{2,1}\}$ defined above, are linearly independent. The following constraint expresses this fact,
\begin{equation}
z^a \chi_a =0.
\end{equation}
The Yukawa couplings in the complex structure sector is defined by
\begin{equation}
\kappa_{\a\b\g}= - \int_{\M} \Om\w \chi_\a^\mu \w \chi_\b^\nu \w \chi_\g^\rho \; \Om_{\mu\nu\rho} \label{eq:comyuk} 
\end{equation}
where
\begin{equation}
\chi_\a^{\mu} = \frac{1}{2\|\Om\|^2} \ol{\Om}^{\mu\rho\sigma} \chi_{\a\; \rho\sigma\bar{\nu}} \; d\xi^{\bar{\nu}}.
\end{equation}
The natural metric (the Weil-Petersson-De Witt metric) on the complex structure moduli space is defined by \cite{candelas,bodner}:
\begin{equation}
G_{\a\bar{\b}} = -\frac{\int_{\M} \chi_\a \w \bar{\chi}_{\bar{\b}}}{\int_{\M} \Om \w \overline{\Om}}.
\end{equation}
In the gauge-choice that we are working in this becomes
\begin{equation}
\int_{\M} \chi_\a \w \bar{\chi}_{\bar{\b}} = 8i \V G_{\a\bar{\b}} \label{eq:cmplxmetric}
\end{equation}
%
\subsubsection{The Large Complex Structure Limit}\label{ap:cmplxlarge}
In Section \ref{sec:defCY} we extracted, from the the expressions of $d\Om$ and $dJ$ on $\tM$ \cite{gurrieri}, explicit expressions for the three non-vanishing torsion classes. To do so we have relied heavily on the assumption that $\tM$ is the `deformation' of some underlying Calabi-Yau manifold $\M$ in the large complex structure. Mirror symmetry for Calabi-Yau manifolds tells us that the prepotential for the complex modulus space of $\M$ in this limit is, in some suitable basis, the same as the bare prepotential of the K\"ahler modulus space of $\W$ in the large volume limit. We can thus set, in the large complex structure limit, the prepotential for $\M$ to
\begin{equation}
\mathscr{G}= -\frac{1}{3!} \frac{\kappa_{\a\b\g}z^\a z^\b z^\g}{z^0} + \frac{1}{2} S_{\a\b}z^\a z^\b + z^0 A_\a z^\a. 
\end{equation}
Since the coordinates $z^a$ are projective we can set $z^0=1$ (this is the gauge choice of \cite{gurrieri}) with the understanding that we do so \emph{after} taking derivatives. The period matrix (see \cite{candelas} for more details) is given by
\begin{equation}
\mathbb{G}_{ab}=\left[\frac{\partial^2 \mathscr{G}}{\partial z^a \partial z^b}\right].
\end{equation}
For our purposes we shall only need the imaginary part of this matrix, which in the large complex structure is given by,
\begin{equation}
\mathbb{G}^- = \left[ \begin{array}{cc}
x^T \kappa x - 2 \V & -(\kappa x)_\a \\
-(\kappa x)_\a & \kappa_{\a \b} \end{array}\right], \label{eq:period}
\end{equation}
where we have introduced real and imaginary parts for the unfixed moduli $z^\a$ by
\begin{equation}
z^\a = x^\a +i y^\a,
\end{equation}
as well as the symmetric $h^{2,1}\times h^{2,1}$ matrix
\begin{equation}
\kappa_{\a\b} = \kappa_{\a\b\g} y^\g
\end{equation}
and identified
\begin{equation}
\V= \frac{1}{3!} \kappa_{\a\b\g} y^\a y^\b y^\g. \label{eq:gaugevol}
\end{equation}
For the last expression we have used the following fact, valid for any $SU(3)$ six-manifold,
\begin{equation}
\Om \w \bar{\Om} = -\frac{4i}{3}  J\w J \w J \label{eq:singlets}
\end{equation}
integrated over $\M$ in the large complex structure limit. The inverse of $\mathbb{G}^-$ is then easily found to be
\begin{equation}
[\mathbb{G}^-]^{-1} =-\frac{1}{2\V} \left[ \begin{array}{cc}
1 & x_\a \\
x_\a & (x^T \kappa x - 2\V) \kappa^{-1}_{\a\b} \end{array} \right].
\end{equation}
In general one can then express the $\b^a$ as
\begin{equation}
\b^a = \frac{i}{2} \left( [\mathbb{G}^-]^{-1}\right)^{ab} \left( k_b \Om - \bar{k}_b \bar{\Om} + \chi_b - \bar{\chi}_{\bar{b}}\right).
\end{equation}
It can also be shown that 
\begin{equation}
\begin{split}
k_\a &= -i \frac{(\kappa y )_\a}{4\V}.
\end{split}
\end{equation}
In the large complex structure limit we then have, with $z^0=1$,
\begin{equation}
\b^0 = -\frac{1}{4\V} \Om^- - \frac{1}{4\V} y^\a \left( \chi_\a + \bar{\chi}_{\bar{\a}} \right). 
\end{equation}
This expression lets us read off the values of $W^+_1$ and $W_3$ in the main text.

We have also used in our computations the following real form of the metric (\ref{eq:cmplxmetric}) on the complex structure moduli space:
\begin{equation}
G_{\a \bar{\b}}= -\frac{1}{4} \frac{\kappa_{\a\b}}{\V} + \frac{(\kappa y)_\a (\kappa y)_\b}{\V^2} \label{eq:largecmplxmetric}
\end{equation}  
%
\section{Derivations of Important Identities} \label{ap:details}
In this appendix we derive the formulae which play a crucial role in the calculations of Section \ref{sec:defCY} of this paper. All of the following expressions are valid in the large complex structure and large volume limit of $\tm$ and are given in the specific gauge $z^0=1$. 
\subsection{Variation of the Integral of Ricci Scalar}
Here, we derive the expression (\ref{eq:varricci}). From the variation of (\ref{eq:intricci}) we have
\begin{equation}
\delta \int_{\tm} d^6y \sqrt{g} R = \delta \left[ \frac{e^2}{8\V}\right]. \label{eq:intriccivarapp}
\end{equation}
First, consider the left hand side of this equation:
\begin{equation}
\begin{split}
\delta \int_{\tm} d^6y \sqrt{g} R &= \int_{\tm} d^6y \left[ (\delta \sqrt{g}) R + \sqrt{g} \delta R\right]\\
&= \int_{\tm}d^6y \sqrt{g}\left[ \frac{1}{2}  g^{mn} \delta g_{mn} R + \delta g^{mn} R_{mn} \right]
\end{split}
\end{equation}  
where we have discarded a term involving $\delta R_{mn}$ since it is a total derivative on $\tm$. Using Strominger's formula (\ref{eq:strominger2}) it is easy to show that 
\begin{equation}
g^{mn} \delta g_{mn} = 8 (v\cdot \delta v).
\end{equation}
On the other hand the right hand side of (\ref{eq:intriccivarapp}) gives
\begin{equation}
\frac{1}{8\V} \delta e^2 - \frac{e^2}{2\V} (v\cdot \delta v)
\end{equation} 
where we have used
\begin{equation}
\delta \frac{1}{\V} = -\frac{4}{\V} (v\cdot \delta v)
\end{equation}
which, in turn, can easily be derived from the definition (\ref{eq:voldef}) of $\V$  and (\ref{eq:strominger2}). Putting all of this together in (\ref{eq:intriccivarapp}) and using (\ref{eq:intricci}) we obtain equation (\ref{eq:varricci}):
\begin{equation}
\int_{\tm} d^6y \delta g^{mn} R_{mn} = \frac{1}{8\V} \delta e^2 - \frac{e^2}{\V} (v\cdot \delta v).
\end{equation}
\subsection{Useful Identities}
In this subsection we derive several identities which are useful for the derivations presented in Section \ref{sec:defCY}.

From eq. (\ref{eq:moduli}) we have
\begin{equation}
\frac{\delta g_{\mu\nu}}{\delta {\bar{z}}^{\bar{\alpha}}} =  \frac{1}{8} (\Om\cdot {\bar{\chi}}_{\bar{\alpha}})_{\mu\nu} 
\end{equation}
This implies that the right hand side is symmetric in $\mu$ and $\nu$.  We then consider
\begin{equation} 
\begin{split}
(\ol{\Om}\cdot \chi_\a)^{\mu\rho} (\Om\cdot \bar{\chi}_{\bar{\b}} )_{\rho\nu}=(\ol{\Om}\cdot \chi_\a)^{\mu\rho} (\Om\cdot \bar{\chi}_{\bar{\b}} )_{\nu\rho}.
\end{split}
\end{equation}
We now use the identities (\ref{eq:projident}) on both sides of this equation to eliminate the $\Om$'s as well as the fact that $\chi$ and $\bar{\chi}$ are primitive. Thus we get
\begin{equation}
2 (\chi_\a)^\mu {}_{\d \tau} (\bar{\chi}_{\bar{\b}})_\nu {}^{\d \tau} = \d^\mu_\nu (\chi_\a)^\rho {}_{\theta\d} (\bar{\chi}_{\bar{\b}})_\rho {}^{\theta \d} - (\chi_\a)^\rho {}_{\nu\d } (\bar{\chi}_{\bar{\beta}})_{\rho} {}^{\mu\d}
\label{eq:usefulident}.
\end{equation}
Similarly by considering $(\ol{\Om}\cdot \chi_\a)^{\mu\rho} (\Om\cdot \bar{\chi}_{\bar{\b}})_{\nu\rho}=(\ol{\Om}\cdot \chi_\a)^{\rho\mu} (\Om\cdot \bar{\chi}_{\bar{\b}})_{\nu\rho}$ it is easy to show that
\begin{equation}
2 (\bar{\chi}_{\bar{\b}})_{\bar{\mu}\rho}{}^\chi (\chi_\a)_\nu {}^\rho {}_\chi = g_{\bar{\mu} \nu} (\chi_\a)^\rho {}_{\psi\chi} (\bar{\chi}_{\bar{\b}})_\rho {}^{\psi\chi} - (\chi_\a)_{\bar{\mu}\psi\chi} (\bar{\chi}_{\bar{\b}})_\nu {}^{\psi\chi}.\label{eq:usefulident2}
\end{equation}
These identities, in conjunction with (\ref{eq:strominger}) or (\ref{eq:strominger2}), help us to disentangle integrals of terms which involve contractions of forms from complex structure and K\"ahler structure moduli spaces.

Next we present another identity used in deriving eqs.\ (\ref{eq:infcond}) and (\ref{eq:intcond}). We consider the variation of $e^2= e_i e_j G^{ij}$, where $G^{ij}$ is the inverse metric on the CY K\"ahler moduli space given by (\ref{eq:invkahler}):
\begin{equation}\begin{split}
\d e^2 &= -e^i e^j \d G_{ij}\\
&= \frac{3}{2} \frac{e^i e^j \mathcal{K}_{ijk} \delta v^k} {\mathcal{K}} - \frac{9}{2} (e^i e^j \mathcal{K}_{ij}) \frac{(\mathcal{K}_i \d v^i)}{\mathcal{K}^2} 
- \frac{9}{2} \frac{(e^i \mathcal{K}_i) (e^j \d \mathcal{K}_j)}{\mathcal{K}^2} + \frac{27}{2} \frac{(\mathcal{K}_i e^i)^2}{\mathcal{K}^3} (\mathcal{K}_i \d v^i).
\end{split}
\end{equation}
We now use (\ref{eq:strominger2}) and the following form of (\ref{eq:modkahlermetric})
\begin{equation}
e^i \d v^j \mathcal{K}_{ij} = - 4 \V (e\cdot \d v) + 16 \V (e\cdot v) (v\cdot \d v)
\end{equation}
to get
\begin{equation}\begin{split}
\d e^2= \frac{1}{4} \frac{e^i e^j \mathcal{K}_{ijk} \d v^k}{\V} - 16 (e\cdot v)^2 (v\cdot \d v) + 4 e^2 (v\cdot \d v) + 8 (e\cdot v) (e\cdot \d v) \label{eq:anotherident}.
\end{split}
\end{equation}
This is used in Section \ref{sec:defCY} to eliminate the $\mathcal{K}_{ijk} e^i e^j \d v^k$ term in (\ref{eq:Xxpression}).

\subsection{Integrals} \label{ap:integrals}
In this subsection we outline the necessary steps required to evaluate the terms $X$ and $Y$ defined by (\ref{eq:intX}) and (\ref{eq:cmplxvar}), respectively.
\subsubsection{The Integral in $X$}
First we evaluate the integral in eq.\ (\ref{eq:intX}):
\begin{equation}\begin{split}
&\frac{1}{16} \frac{(e\cdot v)^2}{\V^2} \int_{\tm} d^6\xi \delta v^i \left[ \frac{1}{2}(\om_i)^{\mu{\bar{\nu}}} \bar{\chi}_{\mu \bar{\sigma} \bar{\tau}} \chi_{\bar{\nu}}{}^{\bar{\sigma} \bar{\tau}} - (\om_i)^{\mu\bar{\nu}} \chi_{\mu \sigma \bar{\tau}} \bar{\chi}_{\bar{\nu}}{}^{\sigma \bar{\tau}} \right]\\
=&\frac{1}{16} \frac{(e\cdot v)^2}{\V^2} \int_{\tm} d^6\xi \delta v^i \left[ (\om_i)^{\mu{\bar{\nu}}} \bar{\chi}_{\mu \bar{\sigma} \bar{\tau}} \chi_{\bar{\nu}}{}^{\bar{\sigma} \bar{\tau}} - (\om_i)^\mu {}_\mu \chi_{\nu \sigma \bar{\tau}} \bar{\chi} {}^{\nu\sigma \bar{\tau}} \right] .
\end{split}
\end{equation}
Where we have used (\ref{eq:usefulident2}) to go from the first line to the second. The $(\om^i)^\mu{}_\mu$ factor in the second term can be brought out of the integral using (\ref{eq:strominger}) and the rest of this term can be integrated to give a factor of the volume (in our gauge) using (\ref{eq:cmplxmetric}), (\ref{eq:gaugevol}) and (\ref{eq:largecmplxmetric}). 
\par
To evaluate the first term above we note that since all six forms are proportional we can write
\begin{equation}
\frac{i}{8}\int (\om_i)^{\mu\bar{\nu}} \chi_{\bar{\nu} \psi\chi} \bar{\chi}_\mu {}^{\psi\chi} d^6 \xi = \int f(y) \;\om_i \w J\w J
\end{equation}
where $f(y)$ is in principle a non-trivial function on $\tm$, which turns out to be a constant. Contracting with $v^i$ and using $J_{\mu\bar{\nu}}= v^i (\om_i)_{\mu\bar{\nu}}$ we find
 \begin{equation}
\int (\om_i)^{\mu\bar{\nu}} \chi_{\bar{\nu} \psi\chi} \bar{\chi}_\mu {}^{\psi\chi} d^6 \xi= -2 \mathcal{K}_i.
\end{equation}
Putting all of this together we have for the two terms:
\begin{equation}
\frac{1}{16} \frac{(e\cdot v)^2}{\V^2} \int_{\tm} d^6\xi \delta v^i \left[ \frac{1}{2}(\om_i)^{\mu{\bar{\nu}}} \bar{\chi}_{\mu \bar{\sigma} \bar{\tau}} \chi_{\bar{\nu}}{}^{\bar{\sigma} \bar{\tau}} - (\om_i)^{\mu\bar{\nu}} \chi_{\mu \sigma \bar{\tau}} \bar{\chi}_{\bar{\nu}}{}^{\sigma \bar{\tau}} \right]=\frac{1}{2} \frac{(e\cdot v)^2}{\V} (v\cdot \delta v).
\end{equation}
\subsubsection{The Integral $Y$}
Next we outline the steps involved in evaluating the integral (\ref{eq:cmplxvar}). It is enough just to consider the integral of $\delta g^{\mu\nu} R_{\mu\nu}$ since the other term follows by complex conjugation. Using (\ref{eq:pure}) and (\ref{eq:moduli}) we have
\begin{equation}\begin{split}
& \int_{\tm} \delta g^{\mu\nu} R_{\mu\nu} \sqrt{g} d^6y = \frac{e^2}{32 \V^2} \delta z^\alpha y^{\bar{\beta}} \int_{\tm} \chi_\alpha \w \bar{\chi}{}_{\bar{\beta}} \\
& - \frac{i}{256} \frac{(e\cdot v)^2}{\V^2} \delta z^\a y^\b y^\g \int_{\tm} \sqrt{g} d^6 \xi (\bar{\Om}\cdot \chi_\a)^{\mu\nu} (\chi_\beta)_\mu {}^{\psi\bar{\sigma}} (\chi_\g)_{\nu \psi \bar{\sigma}} \\
&-  i\frac{e\cdot v}{1024} \delta z^\a \int_{\tm} d^6\xi \sqrt{g} (\bar{\Om} \cdot \chi_\a)^{\mu\nu} \left\{ \frac{1}{2} (e\cdot \om)_\mu {}^\rho (\bar{\chi} \cdot \Om)_{\nu\rho}  +(e\cdot \om )_\theta{}^\rho \bar{\chi}_\mu {}^{\psi \theta} \Om_{\nu\psi\rho}\right\}
\end{split}
\end{equation}
The first term on the right hand side above is evaluated using (\ref{eq:cmplxmetric}) and (\ref{eq:largecmplxmetric}). The second term is an integral that involves a cube of $\chi$ and can be shown to be given essentially by the Yukawa couplings on the complex structure moduli space. This term can then be transformed by using
\begin{equation}
8\V G_{\a\bar{\b}} y^\b = \kappa_{\a\b\g}y^\b y^\g,
\end{equation}
which follows easily from (\ref{eq:largecmplxmetric}). To e\-val\-uate the third and the fourth terms we use the identity (\ref{eq:usefulident}) and find that in both terms only $(\om_i)_\mu {}^\mu$ appears which, using Strominger's formula, can be factored out. The integrals in the third and the fourth terms then become essentially the same as the first term. This way we find
\begin{equation}
\int_{\tm} \delta g^{\mu\nu} R_{\mu\nu} \sqrt{g} d^6y = \frac{i e^2}{4\V} \delta z^\a G_{\a\bar{\b}} y^\b
\end{equation}
Eq. (\ref{eq:cmplxvar}) follows from this.
%

\end{document}